\documentclass[12pt]{article}
\usepackage{a4wide}
\usepackage{amsmath,amssymb,amsfonts,amsthm,amscd}
\usepackage[dvips]{epsfig} 
\usepackage{cite}
\def\be{\begin{equation}}
\def\ee{\end{equation}}

\def\be{\begin{equation}}
\def\ee{\end{equation}}
\def\gs{\mathrel{
   \rlap{\raise 0.511ex \hbox{$>$}}{\lower 0.511ex \hbox{$\sim$}}}}
\def\ls{\mathrel{
   \rlap{\raise 0.511ex \hbox{$<$}}{\lower 0.511ex \hbox{$\sim$}}}}

\newcommand{\ba}{\begin{array}{c}}  
\newcommand{\bad}{\begin{array}{ccc}}
\newcommand{\bea}{\begin{equation} \begin{array}{c}}
\newcommand{\eea}{ \end{array} \end{equation}}
\newcommand{\ea}{\end{array}}

\newcommand{\lsim}{{\;\raise0.3ex\hbox{$<$\kern-0.75em\raise-1.1ex\hbox{$\sim$}} 
\;}} 
\newcommand{\gsim}{{\;\raise0.3ex\hbox{$>$\kern-0.75em\raise-1.1ex\hbox{$\sim$}} 
\;}}

\newcommand{\bvec}[1]{\ensuremath{\boldsymbol{#1}}}

\usepackage{accents}
\usepackage{rotating}
\DeclareMathSymbol{\vecarrow}{\mathord}{letters}{"7E}
\newlength{\lvech}
\newlength{\lvecw}
\newcommand{\lvec}[1]{\ensuremath{\text{%
    \settoheight{\lvech}{$\vecarrow$}\addtolength{\lvech}{-.047ex}%
    \settowidth{\lvecw}{$\vecarrow$}%
    $\accentset{\hspace{.47\lvecw}%
        \begin{rotate}{180}%
            \makebox[0pt]{\raisebox{-\lvech}[0pt][0pt]{$\vecarrow$}}%
        \end{rotate}%
    }{#1}$}}}
\newlength{\lrvecw}
\newcommand{\lrvec}[1]{\ensuremath{\text{%
    \settowidth{\lrvecw}{$\vec{#1}$}%
    $\vec{#1}$\hspace{-\lrvecw}$\lvec{#1}$}}}

\newlength{\diracchlen} 
\newcommand{\dirac}[1]{ 
  \ensuremath{ 
  \text{$ 
  \settowidth{\diracchlen}{$#1$}
  #1
  \hspace{-0.5\diracchlen}
  \makebox[0pt][c]{$/$}
  \hspace{0.5\diracchlen}
$}}}

\begin{document} 
\thispagestyle{empty}
\rightline{hep-ph/0209135}
\rightline{DO-TH 02/15}

\begin{center}
{\large \bf Polarized $J/\psi$ production from $B$ mesons at the Tevatron}

\medskip
{V. Krey\footnote{krey@zylon.physik.uni-dortmund.de}} and 
\smallskip
{K.R.S. Balaji\footnote{balaji@zylon.physik.uni-dortmund.de}}

{\it  Institut f\"ur Theoretische Physik, Universit\"at Dortmund,
Otto-Hahn-Str.4,\\ D-44221 Dortmund, Germany}

\end{center}
\noindent
\begin{abstract}

\end{abstract} 
 In the framework of NRQCD and parton model, we estimate in detail,
the production cross section for polarized $J/\psi$ from $B$ meson decays.
In order to contrast with data, we also take into account additional 
$J/\psi$ production due to decay of excited charmonium states. 
We calculate the helicity parameter, $\alpha$, and as an application,
we study our results for the Tevatron. This is in contrast to the 
earlier studies which were performed for prompt $J/\psi$ production 
from $p \bar p$ collisions. Our estimates are, for $J/\psi$ from $B$ decays,
$\alpha_{J/\psi} = -0.04\pm 0.06$ and for $B$ decays to $\psi^\prime$,
$\alpha_{\psi^\prime}= -0.03\pm 0.07$. These results have been 
evaluated in the $J/\psi$ transverse momentum
interval, $10 ~\mbox{GeV} \leq k_T \leq 30 ~\mbox{GeV}$. In the limit of
the color singlet model, $\alpha$ shows a direct dependence on the 
Peterson parameter, thereby reflecting the dynamics of the $b$ quark 
hadronization. With Run II of the Tevatron, 
it is expected that the fits for $\alpha$ will improve by about a factor of 
50, leading to better limits on the matrix elements. 
\vspace{0.2cm} 


\newpage 
 
\section{Introduction} 
\label{intro}
The production mechanism of bound states involving a heavy 
quark and anti-quark system can be addressed within non-relativistic quantum 
chromodynamics (NRQCD) \cite{Caswell:1986ui}.
In the earliest attempts, charmonium production was described by the color 
singlet model through processes like $B$ decays $(b \rightarrow c \bar c s)$
\cite{DeGrand:1980wf,Wise:1980tp,Kuhn:1980zb} and gluon-gluon fusion 
$(gg \rightarrow c \bar c g)$ \cite{Chang:1980nn}. We refer 
to \cite{Schuler:1994hy} for a review on these issues. 
However, the color singlet models had several problems, e.g., underestimation
of the hadroproduction of charmonium \cite{Vanttinen:1995sd}, the 
$\psi^\prime$ anomaly \cite{Braaten:1994xb,Roy:1994ie} and infrared divergences
in $P$ wave charmonium production, which later had a resolution based on
factorization results \cite{Bodwin:1995jh,Bodwin:1992ye}. 
These problems suggested the need to advance beyond the color singlet model 
or similar variants like the color evaporation model \cite{Fritzsch:1977ay}. 
In a systematic approach, by including the color octet contributions within 
the NRQCD framework it was shown that these problems could indeed be resolved
to a good accuracy \cite{Beneke:1996yb,Beneke:1996tk,Beneke:1996xg}.

Within NRQCD, which is well designed for separating relativistic from 
non-relativistic scales, Bodwin, Braaten and 
Lepage developed a factorization formalism to calculate quarkonium
decays and production \cite{Bodwin:1995jh}. The formalism allows for a 
systematic calculation of the inclusive cross sections to any order in strong 
coupling $\alpha_s$, and an expansion in $v^2$. Here, $v$ is the 
relative velocity of the quark and antiquark and is inversely 
proportional to the heavy quark mass. As an illustration, following
potential model calculations, for bottonium 
states, ${v \sim 0.1}$ while for charmonium states, ${v \sim 0.3}$ indicating 
a better convergence in the perturbative expansion for heavier quark states
\cite{Quigg:1979vr}.
It is interesting to observe that the theory exhibits a scale hierarchy 
of the type, ${m_Q \gg m_Q v \gg m_Qv^2 \sim \Lambda_{\textrm{QCD}}}$, where $m_Q$ is 
the heavy quark mass. Therefore, it is appropriate to use NRQCD as an 
effective field theory with $v$ as the expansion parameter which is also 
a naturally small scale of the theory. In addition, at leading order
in $v$, NRQCD has a strong correspondence to a $1/m_Q$ expansion as in
heavy quark effective theory. 

A large fraction of the existing literature on NRQCD 
employing the factorization formalism, broadly concentrates on one of the 
following two issues; (i) on prompt charmonium production in hadron
\cite{Leibovich:1997pa,Braaten:1999qk,Tang:1996zp,Beneke:1996yb},
$\gamma p$ and $ep$ \cite{Fleming:1997jx} and in 
$e^+ e^-$ \cite{Chang:1997dw,Yuan:1997sn,Boyd:1998km,Klasen:2001cu} 
collisions or (ii) on
charmonium production in hadronic $B$ decays 
\cite{Bodwin:1992qr,Kniehl:1999vf,Beneke:1999gq}. Prompt
production refers to quarkonium (here charmonium) that is created in 
interactions of the colliding particles or their constituents, while charmonium
production is also possible in weak decays of $B$ mesons, which will be  
the focus of the present work. At the functional level, the
calculations in case (i) have been adopted to phenomenologically extract
NRQCD matrix elements from experimental data on charmonium production. This 
is possible, because these calculations 
incorporate bound state effects of the initial hadronic states, usually in the
framework of the QCD improved parton model (PM). In the case of
semi-inclusive $B$ decays with charmonium final states, the 
ACCMM model \cite{Altarelli:1982kh} and the PM \cite{Palmer:1997wv} 
have been successfully adopted for this description. A central feature of these results
has been to illustrate the importance of color octet elements to accommodate
the observed momentum spectra of $J/\psi$.

Quarkonium polarization provides an additional test of the color
octet production mechanism of NRQCD \cite{Beneke:1997jh}. The polarized 
cross section has been calculated for prompt
$\psi$ production \cite{Leibovich:1997pa} as well as for $J/\psi$ production 
in $b$ quark decays \cite{Fleming:1997pt}. We remark that, in the case of 
$J/\psi$ production from $b$ decays, the calculations do not take into 
account bound state effects, although 
they have been estimated to be significant \cite{Ma:2000bz}. Hence, a 
comparison with data in this case may not be too meaningful, given the 
uncertainties, originating from the negligence of the initial hadron, and the
additional errors due to the non-perturbative NRQCD matrix elements. 
  As a salient prediction, within NRQCD, prompt charmonium production is 
expected to be predominantly in transverse polarization state for large 
transverse momenta $(p_T)$ 
\cite{Beneke:1996yb,Leibovich:1997pa,Braaten:1999qk,Cho:1995ih}; but
this prediction is not in agreement with the CDF data 
\cite{Affolder:2000nn}. We note that the polarization prediction arises from 
the dynamics of massless partons and for large $p_T$, the role of
gluon dynamics is important in prompt quarkonium production,
especially through 
the dominance of gluon fragmentation \cite{Braaten:1993rw}. Furthermore, 
a gluon couples easily to the $^3S_1$ color octet state, which is expected 
to be a dominant spectral state in the prompt $J/\psi$ production mechanism 
at the Tevatron. But correspondingly, the charmonium production at
large transverse momenta (with $p_T \geq 20$ GeV) are not fully probed by 
current experiments. Besides, there are large errors in the polarization 
measurements. Therefore, these features alone preclude any 
possible conclusions on the predictions by NRQCD for polarized charmonium 
production.

On the other hand, at moderate transverse momenta (with $p_T \leq 20$ GeV), 
one can perform the polarization studies for prompt charmonium production to 
make an estimate of the color octet elements and also compare with the 
unpolarized cross sections. This is of particular relevance to the Tevatron
where there are no complications due to higher twist effects 
\cite{Beneke:1997yw}. We refer to \cite{Lee:2002au} for an update on prompt
production of polarized charmonium for the Tevatron. Simultaneously, one can 
also study the charmonium
production which is not prompt and in particular estimate the cross sections 
for polarized production. 

It is the goal of the present work to analyze the polarization
predictions for $J/\psi$ from $B$ meson decays at the Tevatron. Unlike in
the case of prompt production, in this process, we do not expect gluon 
fragmentation as a dominant source for $J/\psi$ production, which led to 
predominantly transverse polarized $J/\psi$. Therefore, our calculation 
can serve as an 
independent probe of NRQCD dynamics for polarized charmonium production, 
besides the existing knowledge from prompt production. We employ the PM 
approach as discussed in 
\cite{Palmer:1997wv} to fold the quark level calculations to arrive at a $B$ 
hadron decay. We calculate the helicity parameter, $\alpha$, from the 
production cross section of the three polarization states of the $J/\psi$. We 
observe that a significant drawback of any such analysis are due to our 
present poor understanding of the relevant NRQCD matrix elements. As an 
outcome of our approach, we
note that in the color singlet model (when the octet elements are set to
zero), the $\alpha$ prediction reduces to the details of bound state
effects of the PM. In other words, $\alpha$ depicts a strong 
dependence on the Peterson fragmentation function which describes the 
Fermi motion of the $b$ quark in the $B$ meson. In some sense, this result is
also to be anticipated simply on grounds that the color singlet model 
predictions depend on the shape and form of the initial state wave function
of the decaying system. With data from Run II of Tevatron, which is 
expected to increase the accuracy by a factor of 50, our analysis may be 
useful to tighten the estimates for the 
matrix elements significantly and make our polarization estimates more 
precise \cite{Anikeev:2001rk}. In addition, in the future, a complete global fit/analysis to
quarkonium production will certainly make the predictions more robust.

Our paper is organized as follows. In the next section, for completeness,
we review the basic ideas of NRQCD pertinent to our calculations. 
In section \ref{b-to-jpsi-x}, we introduce the effective Hamiltonian
which describes the quarkonium production process through free $b$ quark 
decays. Using this formalism, we study the
semi-inclusive decay of a free $b$ quark into $J/\psi$. The short-distance
coefficients and the NRQCD matrix elements are explicitly
calculated and the decay width is presented for $b \to J/\psi(\lambda) + X$.
Here, $\lambda$ denotes one of the three helicity states of $J/\psi$. 
Towards the end of this
section, we also discuss an extension of our calculations to excited 
charmonium states. The bound state effects of the initial $B$ meson, whose 
influence hitherto has been neglected in the calculation, are described in 
section \ref{parton-model}. In this analysis, we use the PM
approach to evaluate the bound state effects. Starting with a short 
introduction to the PM, the restrictions of the model applicability 
and estimates for the semi-inclusive decay rate for a $B$ meson with a 
charmonium final state are presented. This is followed by 
section \ref{section-tevatron}, where we describe 
the application of the results derived so far to the Tevatron and
introduce suitable kinematic variables. In section \ref{bcross}, we 
discuss the production cross section for polarized $J/\psi$ at the
Tevatron. In order to phenomenologically implement the production cross section
for $B$ mesons at the Tevatron, we introduce a simple two-parameter fit
procedure. In section \ref{section-nrqcd-matrix-elements}, we describe the
relevant NRQCD matrix elements which we use for this analysis and also 
discuss the various sources of input errors for our estimates. Following this,
in section \ref{section-numerical-analysis}, we give our detailed numerical
estimates for the polarized cross section and predictions for the 
helicity parameter, $\alpha$. We also discuss the relevance/influence of the 
various theoretical input errors to our predictions. The 
differential cross section for $J/\psi$ and $\psi'$ production from $B$ decays 
and the corresponding polarization parameter $\alpha$ are displayed and 
compared with current experimental data \cite{Affolder:2000nn}. 
Since, the data includes
feed-down channels from excited charmonium states, we account for this in our analysis to derive the 
polarization cross sections. Finally in section \ref{summary}, we conclude 
with a summary of the results
and comment on further possible improvements to the precision of our 
calculation. In the appendix of this paper, we have tabulated all the relevant
matrix elements and their sources and give the values which we use in our analysis.


\section{Basic formalism}
\label{section-nrqcd-approach}
 In \cite{Bodwin:1995jh}, it was shown that effects of the lower
momentum scales of the order, $m_Q \cdot v$, $m_Q \cdot v^2$ and 
$\Lambda_{\textrm{QCD}}$ can be factored into matrix elements that are accessible 
only via non-perturbative techniques or from experiments. On the other hand, 
the short distance contributions that occur on scales larger than the heavy
quark mass, can be
calculated within perturbative QCD. A matching prescription is
required to identify the perturbatively calculated short-distance part
and the non-perturbative NRQCD matrix elements. In the following, we 
recollect this matching procedure for polarized quarkonium production
which we shall later use for our calculation. This is followed by 
a basic description of the matrix elements and their scaling
properties.
\subsection{The matching procedure}
Let us consider an inclusive production of a quarkonium state $H$ with 
momentum $k$ and helicity $\lambda$ via a parton level decay process of the 
type $I \to H(k,\lambda) + X$. The semi-inclusive decay width is given as
\begin{equation}
  \sum\limits_X d\Gamma (I \to H(k,\lambda) + X) = \frac{1}{2 E} \frac{d^3k}{(2
    \pi)^3 2 E_H} \sum\limits_X (2 \pi)^4 \delta^{(4)} (p - k - p_X) |{\cal
    T}_{I \to H(k,\lambda) + X}|^2~,
  \label{2eq1}
\end{equation}
where $E$ and $\bvec{p}$ are energy and momentum of the decaying 
particle, $E_H$ is the energy of the quarkonium and the sum over $X$
includes the phase space integration for the additionally produced particles.
On the other hand, from the NRQCD factorization theorem, the decay width in 
(\ref{2eq1}) can be factorized into short-distance coefficients and 
long-distance matrix elements of local four-quark operators. Formally, 
\begin{equation}
  \sum\limits_X d\Gamma (I \to H(k,\lambda) + X) = \frac{1}{2 E} \frac{d^3k}{(2
    \pi)^3 2 E_H} \sum\limits_{m,n} C_{mn}(p, k) \times \langle {\cal
    O}_{mn}^{H(\lambda)} \rangle~.
  \label{2eq2}
\end{equation}
In (\ref{2eq2}), the short-distance coefficients $C_{mn}$ ($m$ and $n$ 
denote some quantum numbers of the various states) depend only on 
kinematical quantities such as momenta and masses of the involved particles. 
They include effects of distances of the order $1/m_Q$ and smaller, where 
$m_Q$ is the mass of the
quarks from which the quarkonium $H$ is built. The matrix elements,
$\langle {\cal O}_{mn}^{H(\lambda)} \rangle$, are expectation
values of local four-quark operators, sandwiched between vacuum
states, $\langle 0| \ldots | 0 \rangle$ . These cannot be calculated
perturbatively, but are extracted from experiment or from lattice
calculations.

As a passing remark, we note that if the decaying particle is a hadron, then 
the parton level decay width in (\ref{2eq2}) must be folded with a suitable 
distribution function for the parton in the initial hadronic state. In this 
case, the factorization approximation requires the final
quarkonium state to carry a large relative transverse momentum
compared to $\Lambda_{\textrm{QCD}}$ \cite{Bodwin:1995jh,Braaten:1996jt}.

The typical four-quark operators which are related to the long-distance 
matrix elements have the general structure
\begin{equation}
  {\cal O}_{mn}^{H(\lambda)} = \psi^\dag {\cal K'}_m^\dag \chi {\cal
    P}_{H(\lambda)} \chi^\dag {\cal K}_n \psi~,
  \label{2eq3}
\end{equation}
where $\psi$ and $\chi$ are the heavy quark and antiquark
non-relativistic field operators,
respectively, and ${\cal K}_n$ and ${\cal K'}_m^\dag$ are products of spin and
color matrices as well as covariant derivatives. Here, ${\cal
  P}_{H(\lambda)}$ is a projection operator that projects onto the subspace of
states that contains the quarkonium state $H(\lambda)$ and in addition
soft hadronic final states denoted by $S_X$. These soft states are supposed 
to be light, due to the NRQCD cut-off requirement, i.e.~their total energy 
has to be less than the NRQCD ultraviolet cut-off $\Lambda$ to avoid double 
counting. Hence including them, ${\cal  P}_{H(\lambda)}$ is written as 
\begin{equation}
  {\cal P}_{H(\lambda)} = \sum\limits_{S_X} |H(\bvec{k} = 0, \lambda) + S_X\rangle
  \langle H(\bvec{k} = 0,\lambda) + S_X|~.
  \label{2eq4}
\end{equation} 
The matching procedure between the complete theory and the NRQCD expression
requires that the normalization of the mesonic states in both the 
frameworks, i.e.,~in (\ref{2eq1}) and (\ref{2eq2}) be the same. In this
analysis, we follow the relativistic normalization procedure as suggested 
in \cite{Braaten:1996jt}. The matching condition is given as
\begin{equation}
\sum\limits_X (2 \pi)^4 \delta^{(4)} (p - k - p_X) {\cal T}^{\ast}_{I \to
    c\bar{c}' + X} {\cal T}_{I \to c\bar{c} + X} = \sum\limits_{m,n} C_{mn}(p,
    k) \langle \psi^\dag {\cal K'}_m^\dag \chi {\cal P}_{H(\lambda)} \chi^\dag
    {\cal K}_n \psi \rangle~.
    \label{2eq8}
\end{equation}
To carry out the matching procedure explicitly, both the l.h.s.~and
the r.h.s.~of 
(\ref{2eq8}) have to be expanded as a Taylor series in $\bvec{q}$ and
$\bvec{q}'$. The short-distance coefficients can then be simply identified by
an order by order comparison in the coupling constant along with $\bvec{q}$
and $\bvec{q}'$.

\subsection{Expansion and simplification of matrix elements}
As mentioned above, an expansion of the matrix elements is necessary
for matching with the complete theory. In addition, by applying 
the symmetries of NRQCD the 
matrix elements are simplified and expressed in terms of standard matrix
elements. The relative magnitudes of each of the matrix elements can be 
estimated, using velocity-scaling rules which we list here. 

In the case of $J/\psi$, the
independent matrix elements can be determined by simple tensor analysis,
since it is a vector meson with $J = 1$. Therefore, the
helicity label, $\lambda$, transforms like a vector index in a
spherical basis. This corresponds to choosing circular polarization
vectors as the basis vectors for the polarization states of $J/\psi$.
A unitary transformation, given by the matrix 
\begin{equation}
  \epsilon_i^{\lambda} = \left( \begin{array}{ccc} -1/\sqrt{2} &
      -i/\sqrt{2} & 0 \\ 0 & 0 & 1 \\ 1/\sqrt{2} & -i/\sqrt{2} & 0
    \end{array} \right)~,
      \label{2eq13a}
\end{equation}
connects the two basis. Here, $i$ runs from $1$ to $3$, whereas
$\lambda$ takes the values $+1$, $0$ and $-1$. 
In what follows, we list the matrix elements that appear in the
calculation of $\psi$ production in $b$ quark decays. Rotational
symmetry as well as heavy quark spin symmetry allow them to be
expressed in terms of standard matrix elements \cite{Bodwin:1995jh}
with well defined spectral states, $^{2 S + 1} L_J$. Details on the
expansion and reduction of NRQCD matrix elements can be found in
\cite{Braaten:1996jt}.
For the simplest matrix elements without any vector indices one finds
\begin{align}
  \langle \psi^\dag \chi {\cal P}_{J/\psi(\lambda)} \chi^\dag \psi
  \rangle &= \tfrac{4}{3} m_c \langle {\cal O}_1^{J/\psi}(^1 S_0)
  \rangle ~, \label{rot-inv-1s0-singlet} \\
  \langle \psi^\dag T^a \chi {\cal P}_{J/\psi(\lambda)} \chi^\dag
  T^a \psi \rangle &= \tfrac{4}{3} m_c \langle {\cal O}_8^{J/\psi}(^1 S_0)
  \rangle~, \label{1s0-octet}
\end{align}
for the color singlet and octet case, respectively. The factor $4 m_c$
originates from the different normalization of states in
\cite{Bodwin:1995jh} and \cite{Braaten:1996jt}. The remaining
dimension 6 matrix elements can be reduced to
\begin{align}
  \langle \psi^\dag \sigma^i \chi {\cal P}_{J/\psi(\lambda)}
  \chi^\dag \sigma^j \psi \rangle &= \tfrac{4}{3} \epsilon_i^{\lambda \dag}
  \epsilon_j^\lambda m_c \langle {\cal O}_1^{J/\psi}(^3 S_1) \rangle~,
  \label{3s1-singlet-helicity} \\
  \langle \psi^\dag \sigma^i T^a \chi {\cal P}_{J/\psi(\lambda)}
  \chi^\dag \sigma^j T^a \psi \rangle &= \tfrac{4}{3} \epsilon_i^{\lambda \dag}
  \epsilon_j^\lambda m_c \langle {\cal O}_8^{J/\psi}(^3 S_1) \rangle~,
  \label{3s1-octet-helicity}
\end{align}
up to corrections of order $v^2$. In case of the matrix elements
with four vector indices we have
\begin{align}
   \left\langle \psi^\dag \left( - \tfrac{i}{2} \lrvec{D}^k
     \right) \sigma^l \chi {\cal P}_{c\bar{c},c\bar{c}'} \chi^\dag \left( -
       \tfrac{i}{2} \lrvec{D}^n \right) \sigma^p \psi
     \right\rangle &= 4 \epsilon_i^{\lambda \dag} \epsilon_j^\lambda \delta^{kn}
     m_c \langle {\cal O}_1^{J/\psi}(^3 P_0) \rangle~,
   \label{3p0-singlet-helicity} \\
   \left\langle \psi^\dag \left( - \tfrac{i}{2} \lrvec{D}^k
     \right) \sigma^l T^a \chi {\cal P}_{c\bar{c},c\bar{c}'} \chi^\dag \left( -
       \tfrac{i}{2} \lrvec{D}^n \right) \sigma^p T^a \psi
     \right\rangle &= 4 \epsilon_i^{\lambda \dag} \epsilon_j^\lambda \delta^{kn}
     m_c \langle {\cal O}_8^{J/\psi}(^3 P_0) \rangle ~,
   \label{3p0-octet-helicity}
\end{align}
which also receive corrections at order $v^2$. Other useful relations can
be established on the basis of heavy quark spin symmetry. It relates
the matrix elements with the same orbital angular momentum $L$ and
different total angular momentum $J$ to each other. For instance, the
$P$ wave matrix elements at order $v^2$ are equal up to a multiplicity
factor 
\begin{equation}
  \langle {\cal O}_n^{J/\psi}(^3 P_J) \rangle \approx (2 J + 1) \langle {\cal
    O}_n^{J/\psi}(^3 P_0) \rangle ~.
  \label{2eq19}
\end{equation}
Furthermore, the color singlet matrix elements can be related to the
non-relativistic quarkonium wave function, whose radial part is
denoted by $\bar{R}_\psi$, evaluated at the origin. This can be
achieved by means of the vacuum saturation approximation (VSA).
\begin{equation}
  \langle \psi^\dag \sigma^i \chi {\cal P}_{J/\psi(\lambda)} \chi^\dag
  \sigma^j \psi \rangle = \epsilon_i^\lambda \epsilon_j^{\lambda \dag}
  \frac{3}{\pi} M_\psi | \bar{R}_\psi|^2~,
  \label{vsa-singlet-wavefunction}
\end{equation}
where due to the VSA an error of order $O(v^4 m_c | \bar{R}_\psi|^2)$
is induced.

As mentioned earlier, the non-perturbative matrix elements relative
importance is determined according to the velocity scaling rules. These 
velocity-scaling properties for the $J/\psi$ production matrix
elements are summarized in table \ref{table-velocity-scaling}.
\begin{table}[htb]
  \begin{center}
    \begin{tabular}{|l|c|}
      \hline
      matrix element & $v$-scaling \\
      \hline
      $\langle {\cal O}_1^{J/\psi} (^{3} S_1) \rangle$ & $v^3$ \\
      $\langle {\cal O}_8^{J/\psi} (^{3} S_1) \rangle$ & $v^7$ \\
      $\langle {\cal O}_8^{J/\psi} (^{3} P_J) \rangle$ & $v^7$ \\
      $\langle {\cal O}_8^{J/\psi} (^{1} S_0) \rangle$ & $v^7$ \\
      $\langle {\cal O}_1^{J/\psi} (^{3} P_J) \rangle$ & $v^{11}$ \\
      $\langle {\cal O}_1^{J/\psi} (^{1} S_0) \rangle$ & $v^{11}$ \\
      \hline
    \end{tabular}
  \end{center}
  \caption{Summary of velocity scaling rules for NRQCD matrix
    elements}
  \label{table-velocity-scaling}
\end{table}
\section{The process $b \to J/\psi + X$}
\label{b-to-jpsi-x}
\begin{figure}[htb]
  \begin{center}
    \epsfig{file=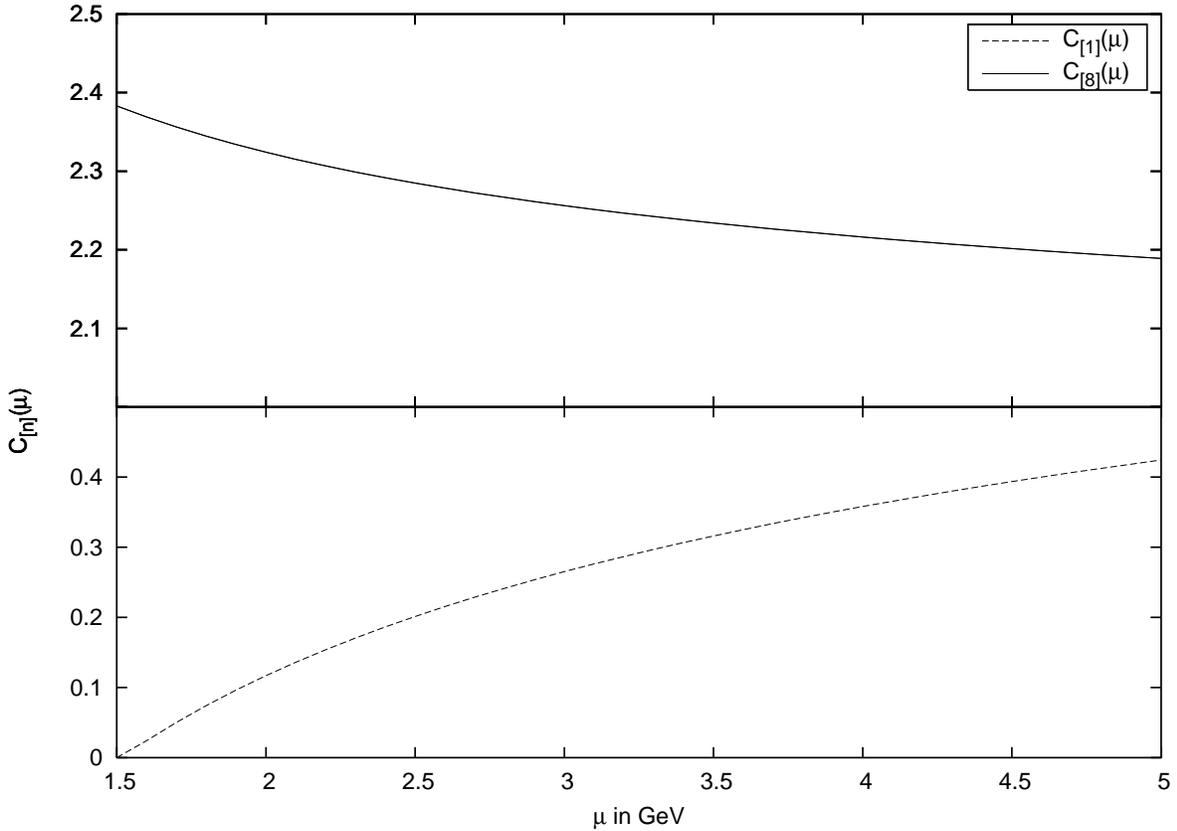,angle=-90,width=16cm}
  \end{center}
  \caption{Dependence of the Wilson coefficients $C_{[n]}(\mu)$ on the
  scale $\mu$.}
  \label{plot-wilson-coefficients}
\end{figure}

Having given the basic 
formalism, we now proceed to describe $J/\psi$ production 
from $b$ decays.
The decay of a $b$ quark is a weakly induced process and is described by the 
exchange of a $W$ boson, transforming the $b$ into a $c$ quark. The $W$ 
subsequently decays into a $\bar{c} q_f$ pair, where the $q_f$ is a light
state and can be either a $s$ or a $d$ quark. At the scale $\mu \sim m_b$, 
after integrating out the $W$ boson, the effective QCD corrected Hamiltonian
\begin{equation}
  {\cal H}_{\rm eff} = \frac{G_F}{\sqrt{2}}
  V_{cb}^{\ast} V_{cf} \left\{ \tfrac{1}{3} C_{[1]}(\mu) \cdot {\cal
  O}_1 + C_{[8]}(\mu) \cdot {\cal O}_8 \right\}
\label{eq-eff-hamiltonian}
\end{equation}
induces the $b \to c \bar{c} + q_f$ transition.
The relevant operators are
\begin{align}
  {\cal O}_1 &= \bar{c} \gamma_\mu L c \bar{q}_f \gamma^\mu L b~, \\
  {\cal O}_8 &=  \bar{c} \gamma_\mu L T^a c \bar{q}_f \gamma^\mu L T^a b~, 
\end{align}
with $L = 1 - \gamma_5$. Here, $C_{[1]}(\mu)$
and $C_{[8]}(\mu)$ are the effective Wilson couplings for the color singlet 
and the color octet operator, respectively. The Wilson coefficients should
not be related with the short-distance coefficients of NRQCD,
denoted by $C_{mn}$. To get a qualitative feeling for the relative strengths 
and their dependence on the factorization scale $\mu$, we have shown the couplings 
in figure \ref{plot-wilson-coefficients} for $\mu \sim m_c \ldots
m_b$. Here, $\alpha_s(M_Z) = 0.119$ has been taken, which corresponds to
$\Lambda_{\textrm{QCD}}^{(n_f = 5)} = 93.14$~MeV at LO. 
In particular, it has been noted that the color singlet coefficient
$C_{[1]}$ exhibits a strong dependence on $\mu$ and even vanishes near
$\mu \sim m_c$ at LO \cite{Bergstrom:1994vc}. This behavior which
hints towards large higher order corrections can not be cured at NLO
\cite{Beneke:1998ks}. We will later allude to this problem, when we discuss 
the various uncertainties pertaining to our results.

\subsection{The decay width}
Applying the effective Hamiltonian (\ref{eq-eff-hamiltonian}) to the
$b \to c \bar{c} + q_f$ decay, we calculate the matrix element, ${\cal T}$ 
at LO to be
\begin{align}
  {\cal T}_{b \to c\bar{c} + q_f} = \frac{G_F}{\sqrt{2}}
  V_{cb}^{\ast} V_{cf} \left\{\tfrac{1}{3} \right. & \left. C_{[1]}
  \cdot \bar{c}(p) \gamma_\mu L c(\bar{p}) \bar{q}_f(p_f) 
    \gamma^\mu L b(p_b) \right. \nonumber \\
  + & \left. C_{[8]} \cdot \bar{c}(p) \gamma_\mu L T^a c(\bar{p})
    \bar{q}_f(p_f) \gamma^\mu L T^a b(p_b) \right\}~. \label{eq-t-matrix}
\end{align}
The four-momenta of the outgoing $c$ and $\bar{c}$ quarks can be expressed as,
${p = \frac{1}{2} k + L\bvec{q}}$ and ${\bar{p} = \frac{1}{2} k - L\bvec{q}}$, 
where $k$ is the total four-momentum of the $c \bar{c}$-system and $\bvec{q}$
is the relative three-momentum in the $c \bar{c}$-rest frame. $L_i^\mu$ is
the Lorentz boost matrix that connects the two frames. Expanding
${\cal T}$ to linear order in $\bvec{q}$, we get
\begin{align}
  {\cal T}_{b \to c\bar{c} + q_f} =& \frac{G_F}{\sqrt{2}}
  V_{cb}^{\ast} V_{cf} \left\{ \tfrac{1}{3} C_{[1]} \cdot
    \bar{q}_f(p_f) \gamma^\mu L b(p_b) \left[2 L_i^\mu \left(m_c \xi^\dag
        \sigma^i \eta + i \epsilon_{ikl} \xi^\dag q^k \sigma^l \eta \right) - k^\mu
      \xi^\dag \eta\right] \right. \nonumber \\
  & \left. + C_{[8]} \cdot \bar{q}_f(p_f) \gamma^\mu L T^a b(p_b)
    \left[2 L_i^\mu \left(m_c \xi^\dag \sigma^i T^a \eta + i \epsilon_{ikl}
        \xi^\dag q^k \sigma^l T^a \eta \right) - k^\mu \xi^\dag T^a \eta\right]
  \right\}~.
\label{eq-t-matrix-exp}
\end{align}
The boost matrices, $L_i^\mu$, also have to be expanded in $\bvec{q}$ and 
to linear order are given to be
\begin{align}
  L_i^0 &= \frac{1}{2 m_c} k_i~, \label{eq-lorentz-exp-a} \\
  L_i^j &= \delta_{ij} + \left(\frac{E_\psi}{2 m_c} - 1\right) \hat{k}_i \hat{k}_j~,
  \label{eq-lorentz-exp-b}
\end{align}
where the hats denote unit vectors.
At LO, $|\bar{\cal T}|^2$ factorizes into a product of two rank two tensors:
\begin{equation}
  |{\cal T}_{b \to c\bar{c} + q_f}|^2 = \tfrac{1}{2} G_F^2 |V_{cb}^{\ast}
  V_{cf}|^2 W^{\mu \nu} \left(T^{(1)}_{\mu \nu} + T^{(8)}_{\mu
  \nu}\right)~. \label{eq-t-matrix-sq}
\end{equation}
$W_{\mu \nu}$ describes the transition of a $b$ quark to $q_f$ and
has the simple form
\begin{equation}
  W_{\mu \nu} = \tfrac{1}{2} \cdot \langle (\dirac{p}_f + m_f) \gamma_\mu
  L (\dirac{p}_b + m_b) \gamma_\nu L \rangle
  = 4 \cdot p_f^\sigma p_b^\lambda \left(S_{\sigma \mu \lambda
      \nu} - i \epsilon_{\sigma \mu \lambda \nu}\right)~,
  \label{eq-w-tensor}
\end{equation}
with
\begin{equation}
  S_{\mu \sigma \nu \lambda} = g_{\mu \sigma} g_{\nu \lambda} - g_{\mu
  \nu} g_{\sigma \lambda} + g_{\mu \lambda} g_{\nu \sigma}~,
\end{equation}
and $\epsilon_{\mu \sigma \nu \lambda}$ being the usual antisymmetric
Levi-Civit\`a symbol.
$T^{(n)}_{\mu \nu}$ refers to $J/\psi$ production in either color
singlet $(n = 1)$ or color octet $(n = 8)$ channel and is obtained as
\begin{align}
  T^{\mu \nu}_{(1)} = \tfrac{1}{9} C_{[1]}^2 \cdot
  & \left[2 L_i^\mu \left(m_c \xi^\dag \sigma^i \eta +
      i \epsilon_{ikl} \xi^\dag q^k \sigma^l \eta \right) - k^\mu \xi^\dag
    \eta\right] \nonumber \\
  \cdot & \left[2 L_j^\nu \left(m_c {\eta'}{^\dag} \sigma^j \xi' +
      i \epsilon_{jnp} {\eta'}{^\dag} {q'}{^n} \sigma^p \xi' \right) - k^\nu {\eta'}{^\dag}
    \xi'\right]~, \label{eq-t-tensor-singlet} \\
  T^{\mu \nu}_{(8)} = \tfrac{1}{6} \delta_{ab} C_{[8]}^2 \cdot
  & \left[2 L_i^\mu \left(m_c \xi^\dag \sigma^i T^a \eta +
          i \epsilon_{ikl} \xi^\dag q^k \sigma^l T^a \eta \right) - k^\mu
    \xi^\dag T^a \eta\right] \nonumber \\
  \cdot & \left[2 L_j^\nu \left(m_c {\eta'}{^\dag} \sigma^j T^b \xi' +
      i \epsilon_{jnp} {\eta'}{^\dag} {q'}{^n} \sigma^p T^b \xi' \right) -
    k^\nu {\eta'}{^\dag} T^b \xi'\right]~. \label{eq-t-tensor-octet}
\end{align}
In the context of the PM, the tensor $W_{\mu \nu}$ will be
replaced by a more general hadronic tensor structure which includes a
distribution function for the heavy $b$ quark. This will be discussed in
section \ref{parton-model}.
In the following, for clarity, we describe in some detail, our calculation 
for obtaining the polarized decay spectrum. 
Contracting the two tensor structures $T_{\mu\nu}^{(n)}$ and
$W_{\mu\nu}$, one can identify six different non-relativistic
four-quark operators which then will be matched to the
corresponding NRQCD operators that have been presented in section
\ref{section-nrqcd-approach}. The contraction can be divided into two
steps, i.e.,~(i) the contraction of the Minkowski indices and (ii) the
contraction of the three-vector indices. In the first step only four
quantities have to be calculated, since $T_{\mu\nu}^{(n)}$ consists of
the structures $L_\mu^i L_\nu^j$ and $k_\mu k_\nu$ (mixed structures like
$L_\mu^i k_\nu$ vanish by symmetry arguments) and $W_{\mu \nu}$
consists of a symmetric part, proportional to $S_{\sigma \mu \lambda
  \nu}$, and an antisymmetric part, proportional to $\epsilon_{\sigma
  \mu \lambda \nu}$. The four quantities  will be denoted by $P$ with a superscript
$(s)$ for symmetric and $(a)$ for antisymmetric, specifying which part of
$W_{\mu \nu}$ they come from. The symmetric terms are
\begin{align}
 P_{ij}^{(s)} &= 4 \cdot p_f^\sigma p_b^\lambda S_{\sigma \mu \lambda \nu} L_i^\mu
 L_j^\nu \nonumber \\
 &= 4 \left(2 p_{b \mu} p_{b \nu} L_i^\mu L_j^\nu - (p_b p_f) g_{\mu \nu}
   L_i^\mu L_j^\nu \right) 
 = 8 p_{b \mu} p_{b \nu} L_i^\mu L_j^\nu + 4 (p_b p_f) \delta_{ij}~,
 \label{eq-p-sym-a} \\
 P^{(s)} &= 4 \cdot p_f^\sigma p_b^\lambda S_{\sigma \mu \lambda \nu} 
k^\mu k^\nu = 8 (p_f k)(p_b k) - 4 (p_b p_f) k^2~. 
\label{eq-p-sym-b} \\
\intertext{Similarly one gets for the antisymmetric part}
 P_{ij}^{(a)} &=  - 4 i \cdot p_f^\sigma p_b^\lambda 
\epsilon_{\sigma \mu \lambda \nu}
  L_i^\mu L_j^\nu = 4 i \epsilon_{\sigma \mu \lambda \nu} k^\sigma p_b^\lambda  L_i^\mu
  L_j^\nu 
 = 4 i p_b^\lambda \sqrt{k^2} \left(- \epsilon_{ijm} L_{\lambda m}\right)~,
  \label{eq-p-asym-a} \\
  P^{(a)} &= - 4 i \cdot p_f^\sigma p_b^\lambda \epsilon_{\sigma \mu \lambda \nu}
  k^\mu k^\nu = 0~. 
\label{eq-p-asym-b}
\end{align}
Following this, we choose a reference frame in which the decaying $b$ quark 
moves with arbitrary three-momentum, $|\bvec{p}_b|$, and corresponding energy, $E_b =
\sqrt{m_b^2 + |\bvec{p}_b|^2}$. The $J/\psi$ three-momentum is denoted by
$|\bvec{k}|$ with energy, $E_\psi = \sqrt{M_\psi^2 + |\bvec{k}|^2}$. The
three-vectors, $\bvec{p}_b$ and $\bvec{k}$, enclose an angle
$\vartheta$. For this choice of reference frame, the above
projectors (\ref{eq-p-sym-a}) - (\ref{eq-p-asym-b}) are evaluated to be
\begin{align}
  P_{ij}^{(s)} &= 8 \bigg[\left(\frac{E_b |\bvec{k}|}{2 m_c} + \left(1 -
      \frac{E_\psi}{2 m_c}\right) |\bvec{p}_b| \cos{\vartheta}\right)^2 \hat{k}_i
  \hat{k}_j - \left(\frac{E_b |\bvec{k}|}{2 m_c} + \left(1 - \frac{E_\psi}{2
        m_c}\right) |\bvec{p}_b| \cos{\vartheta}\right) \nonumber \\
  & \cdot \left(\hat{k}_i p_{b j} + p_{b i} \hat{k}_j\right) + p_{b i} p_{b
    j} \bigg] + 4 \left[m_b^2 - E_b E_\psi + |\bvec{p}_b| |\bvec{k}|
    \cos{\vartheta} \right] \delta_{ij} ~,\label{eq-p-sym-specicifc-a}\\
  P^{(s)} &= 8 \left(E_b E_\psi - |\bvec{p}_b| |\bvec{k}|
    \cos{\vartheta}\right)^2 - 4 \left(m_b^2 + E_b E_\psi
    - |\bvec{p}_b| |\bvec{k}| \cos{\vartheta}\right) (2 m_c)^2 ~,
\label{eq-p-sym-specicifc-b}\\
  P_{ij}^{(a)} &= - 4 i m_c \epsilon_{ijm} \left[ \left(\frac{E_b
        |\bvec{k}|}{2 m_c} + \left(1  - \frac{E_\psi}{2 m_c}\right) 
|\bvec{p}_b|\cos{\vartheta}\right) \hat{k}_m - p_{b m} \right] ~.
\label{eq-p-asym-specific-a}
\end{align}
Collecting together (\ref{eq-w-tensor} -- \ref{eq-t-tensor-octet}) and
(\ref{eq-p-sym-specicifc-a} -- \ref{eq-p-asym-specific-a})
the color singlet contribution is given to be 
\begin{multline}
  W^{\mu \nu} T^{(1)}_{\mu \nu} = \tfrac{1}{9} C_{[1]}^2 \Big\{4
    \left(P_{ij}^{(s)} + P_{ij}^{(a)}\right)
    \left[m_c^2 \xi^\dag \sigma^i \eta {\eta'}{^\dag} \sigma^j \xi'
    \right.  \\
  \left. + \epsilon_{ikl} \epsilon_{jnp} q^k {q'}{^n} \xi^\dag \sigma^l \eta
      {\eta'}{^\dag} \sigma^p \xi'\right] + P^{(s)} \xi^\dag \eta {\eta'}{^\dag}
    \xi' \Big\}~,
  \label{eq-w-t-contract-singlet}
\end{multline}
and for the octet we obtain
\begin{multline}
  W^{\mu \nu} T^{(8)}_{\mu \nu} = \tfrac{1}{6} C_{[8]}^2 \Big\{4
    \left(P_{ij}^{(s)} + P_{ij}^{(a)}\right)
    \left[m_c^2 \xi^\dag \sigma^i T^a \eta {\eta'}{^\dag} \sigma^j T^a \xi'
    \right. \\
  \left. + \epsilon_{ikl} \epsilon_{jnp} q^k {q'}{^n} \xi^\dag \sigma^l T^a \eta
      {\eta'}{^\dag} \sigma^p T^a \xi'\right] + P^{(s)} \xi^\dag T^a \eta {\eta'}{^\dag}
    T^a \xi' \Big\}~.
  \label{eq-w-t-contract-octet}
\end{multline}
In order to perform the matching procedure described earlier, we need to
insert the above results into (\ref{eq-t-matrix-sq}) to get the
squared matrix element, $|{\cal T}_{b \to c \bar{c} + q_f}|^2$.
Following the results of section \ref{section-nrqcd-approach}, we identify 
the short-distance coefficients $C_{mn}$ by making use of the matching 
condition (\ref{2eq8}). As stated before, the integration 
over the phase space of the additionally produced particles (which in our 
case is the $q_f$ quark) has to be included into the
sum over the hadronic rest $X$ on the l.h.s. of (\ref{2eq8}). Using the
standard identity
\begin{equation}
  \int \frac{d^3 p_f}{(2 \pi)^3 2 E_f} = \int \frac{d^4 p_f}{(2 \pi)^3}
  \delta(p_f^2 - m_f^2) \theta(p_f^0)~,
\end{equation}
and performing the four-dimensional phase space integral over $d^4 p_f$, due
to the presence of the $\delta^{(4)}(p_b - k - p_f)$ function, $p_f$ gets 
replaced by $p_b - k$. Thus, the matching condition is obtained to be
\begin{equation}
  \sum\limits_{q_f} 2 \pi \delta [(p_b - k)^2 - m_f^2] \theta [(p_b -
    k)_0] {\cal T}^{\ast}_{b \to c\bar{c}' + q_f} {\cal T}_{b \to
    c\bar{c} + q_f} = \sum\limits_{m,n} C_{mn}(p_b, k) \langle
    \psi^\dag {\cal K'}_m^\dag \chi {\cal P}_{J/\psi(\lambda)}
    \chi^\dag {\cal K}_n \psi \rangle~.
    \label{eq-match-modified}
\end{equation}
We begin with the color octet contributions whose short-distance
coefficients we identify for the spectral states $^{2S + 1} L_J$ of
the $c \bar{c}$ pair. In the following, the three-vector indices on
the l.h.s.~have been suppressed.
\begin{multline}
  C_8[^1 S_0] = \frac{2 \pi}{3} G_F^2 |V_{cb}^{\ast} V_{cf}|^2
  C_{[8]}^2 \delta ((p_b - k)^2 - m_f^2) \\
   \times \left\{ \left( - E_b E_\psi + |\bvec{p}_b| |\bvec{k}|
      \cos{\vartheta} \right) + \frac{1}{2 m_c^2} \left( E_b E_\psi - |\bvec{p}_b|
      |\bvec{k}| \cos{\vartheta} \right)^2 - m_b^2 \right\}~.
  \label{eq-c-1s0-octet}
\end{multline}
\begin{multline}
  C_8[^3 S_1] = \frac{2 \pi}{3} G_F^2 |V_{cb}^{\ast} V_{cf}|^2
  C_{[8]}^2 \delta ((p_b - k)^2 - m_f^2) \\
  \times \left\{ 2 \left[R(|\bvec{p}_b|, |\bvec{k}|, \cos{\vartheta})
  \hat{k}_{i} - {p}_{bi}\right] \left[R(|\bvec{p}_b|, |\bvec{k}|,
  \cos{\vartheta}) \hat{k}_{j} - {p}_{bj}\right] \right. \\
   \left. + \left(m_b^2 - E_b E_\psi + |\bvec{p}_b| |\bvec{k}| \cos{\vartheta} \right)
  \delta_{ij} - 2 m_c i \left[ R(|\bvec{p}_b|, |\bvec{k}|, \cos{\vartheta})
  \hat{k}_{m} - {p}_{bm} \right] \epsilon_{ijm} \right\}~.
  \label{eq-c-3s1-octet}
\end{multline}
\begin{multline}
  C_8[^3 P_0] = \frac{2 \pi}{3} G_F^2 |V_{cb}^{\ast} V_{cf}|^2
  C_{[8]}^2 \delta ((p_b - k)^2 - m_f^2) \\
  \times \Big\{ 2 \left[R(|\bvec{p}_b|, |\bvec{k}|, \cos{\vartheta})
  \hat{k}_{i} - {p}_{bi} \right] \left[R(|\bvec{p}_b|, |\bvec{k}|,
  \cos{\vartheta}) \hat{k}_{j} - {p}_{bj}\right] \epsilon_{ikl} \epsilon_{jnp} \\
   + \left(m_b^2 - E_b E_\psi + |\bvec{p}_b| |\bvec{k}| \cos{\vartheta} \right)
  \left(\delta_{kn} \delta_{lp} - \delta_{kp} \delta_{ln}\right) \\ 
- 2 m_c i \Big[ R(|\bvec{p}_b|, |\bvec{k}|, \cos{\vartheta}) \hat{k}_m 
   - {p}_{bm} \Big]\left( \delta_{ml}
  \epsilon_{knp} - \delta_{mk} \epsilon_{lnp} \right) \Big\} \frac{1}{m_c^2}~,
  \label{eq-c-3p0-octet}
\end{multline}
where the kinematic function
\begin{equation}
  R(|\bvec{p}_b|, |\bvec{k}|, \cos{\vartheta})=\frac{E_b |\bvec{k}|}{2 m_c} +
    \left(1 - \frac{E_\psi}{2 m_c}\right) |\bvec{p}_b| \cos{\vartheta}~.
  \label{eq-kin-abb-1}
\end{equation}
The color singlet short-distance coefficients $C_1[^{2S + 1} L_J]$ is most easily
obtained from the corresponding octet coefficient by replacing the
color matrices $T^a$ by unit matrices as well as changing the Wilson
coefficient from $C_{[8]}$ to $C_{[1]}$ along with an
overall factor of $\frac{2}{3}$. This also serves as a useful book-keeping
device for our calculations.

>From table \ref{table-velocity-scaling}, it is seen that the 
color singlet matrix elements with angular quantum numbers $^1 S_0$ and $^3
P_0$ scale with $v^{8}$ relative to the baseline matrix
element. Additionally, the color singlet production is suppressed
relative to the color octet production. This follows from the
comparison of their Wilson coefficients, whose squared ratio turns out
to be $C_{[8]}^2/C_{[1]}^2 \sim 25$, which can be estimated from figure
\ref{plot-wilson-coefficients}. Therefore, the contributions of 
these color singlet matrix elements are highly suppressed and one 
only needs to take into account the $^3 S_1$ contribution for the
singlet case. On the other hand, all the three octet matrix elements should
be included, because they scale as $v^4$ relative to the dominant
color singlet $^3 S_1$ matrix element and are enhanced due to the larger 
Wilson coefficient $C_{[8]}$. Thus, we have the singlet contribution
\begin{multline}
  C_1[^3 S_1] = 4 \pi G_F^2 |V_{cb}^{\ast} V_{cf}|^2
  \tfrac{1}{9} C_{[1]}^2 \delta ((p_b - k)^2 - m_f^2) \\
  \times \left\{ 2 \left[R(|\bvec{p}_b|, |\bvec{k}|, \cos{\vartheta})
  \hat{k}_{i} - {p}_{bi}\right] \left[R(|\bvec{p}_b|, |\bvec{k}|,
  \cos{\vartheta}) \hat{k}_{j} - {p}_{bj}\right] \right. \\
   \left. + \left(m_b^2 - E_b E_\psi + |\bvec{p}_b| |\bvec{k}| \cos{\vartheta} \right)
  \delta_{ij} - 2 m_c i \left[ R(|\bvec{p}_b|, |\bvec{k}|, \cos{\vartheta})
  \hat{k}_{m} - {p}_{bm} \right] \epsilon_{ijm} \right\}~.
  \label{eq-c-3s1-singlet}
\end{multline}
In order to calculate the decay rate, we choose a reference frame where the 
$J/\psi$ moves along the positive $z$ axis. The $b$ quark momentum vector is 
then most conveniently parameterized in spherical polar coordinates, where 
$\vartheta$ is the polar and $\varphi$ is the azimuthal angle. We 
have the unit vectors
\begin{align}
  \hat{k}_{i} &= \delta_{i3}~,   \label{3eq15a} \\
  \hat{p}_{bi} &= \cos{\varphi} \sin{\vartheta} \delta_{i1} + \sin{\varphi}
    \sin{\vartheta} \delta_{i2} + \cos{\vartheta} \delta_{i3}~.  \label{3eq15b}
\end{align}
Multiplying the short-distance coefficients $C_n[^{2S + 1} L_J]$ from
(\ref{eq-c-1s0-octet} - \ref{eq-c-3s1-singlet}) along with the
appropriate matrix elements, we get the 
differential decay rate as in (\ref{2eq2}). After contracting the
short- with the long-distance part we are left with a triple
differential decay width for a $b$ quark, moving at arbitrary
momentum $|\bvec{p}_b|$ that decays into a $J/\psi$ with helicity
$\lambda$. The individual matrix element contributions in this case are 
\begin{multline}
  E_\psi E_b \frac{d^3 \Gamma_1[^3 S_1]}{d k^3} =  \frac{1}{6 \pi^2}
  G_F^2 |V_{cb}^{\ast} V_{cf}|^2  \tfrac{1}{9} C_{[1]}^2 m_c
  \delta ((p_b - k)^2 - m_f^2) \\
  \Big\{ \delta_{\lambda 0} \left[ \tfrac{1}{2 m_c^2} \left( E_b |\bvec{k}|
      - E_\psi |\bvec{p}_b| \cos{\vartheta} \right)^2 - |\bvec{p}_b|^2
    (1 - \cos^2{\vartheta}) \right] + \lambda \left[ -E_b |\bvec{k}|
    + E_\psi |\bvec{p}_b| \cos{\vartheta} \right] \\
  + 1 \left[ m_b^2 + |\bvec{p}_b|^2 (1 - \cos^2{\vartheta}) - E_b
    E_\psi + |\bvec{p}_b| |\bvec{k}|
    \cos{\vartheta} \right] \Big\} \times \langle {\cal
  O}_1^{J/\psi}(^3 S_1) \rangle~,
  \label{ddGamma-3s1-singlet}
\end{multline}
\begin{multline}
  E_\psi E_b \frac{d^3 \Gamma_8[^1 S_0]}{d k^3} =  \frac{1}{36 \pi^2}
  G_F^2 |V_{cb}^{\ast} V_{cf}|^2  C_{[8]}^2 m_c \delta ((p_b - k)^2 - m_f^2)
  \\
   \Big\{ 1 \left[ -E_b E_\psi + |\bvec{p}_b| |\bvec{k}| \cos{\vartheta} -
    m_b^2 + \tfrac{1}{2 m_c^2} \left( E_b E_\psi - |\bvec{p}_b| |\bvec{k}|
      \cos{\vartheta} \right)^2 \right] \Big\} \times \langle {\cal
  O}_8^{J/\psi}(^1 S_0) \rangle~,
  \label{ddGamma-1s0-octet}
\end{multline}
\begin{multline}
  E_\psi E_b \frac{d^3 \Gamma_8[^3 S_1]}{d k^3} =  \frac{1}{36 \pi^2}
  G_F^2 |V_{cb}^{\ast} V_{cf}|^2  C_{[8]}^2 m_c \delta ((p_b - k)^2 - m_f^2) \\
  \Big\{ \delta_{\lambda 0} \left[ \tfrac{1}{2 m_c^2} \left( E_b |\bvec{k}|
      - E_\psi |\bvec{p}_b| \cos{\vartheta} \right)^2 - |\bvec{p}_b|^2
    (1 - \cos^2{\vartheta}) \right] + \lambda \left[ -E_b |\bvec{k}|
    + E_\psi |\bvec{p}_b| \cos{\vartheta} \right] \\
  + 1 \left[ m_b^2 + |\bvec{p}_b|^2 (1 - \cos^2{\vartheta}) - E_b
    E_\psi + |\bvec{p}_b| |\bvec{k}| \cos{\vartheta} \right] \Big\}
  \times \langle {\cal O}_8^{J/\psi}(^3 S_1) \rangle~,
  \label{ddGamma-3s1-octet}
\end{multline}
\begin{multline}
  E_\psi E_b \frac{d^3 \Gamma_8[^3 P_0]}{d k^3} = \frac{1}{12 \pi^2}
  G_F^2 |V_{cb}^{\ast} V_{cf}|^2  C_{[8]}^2 m_c \delta ((p_b - k)^2 - m_f^2) \\
  \Big\{ \delta_{\lambda 0} \left[ \tfrac{1}{2} |\bvec{p}_b|^2 (1 -
    \cos^2{\vartheta}) - \tfrac{1}{2 m_c^2} \left(E_b |\bvec{k}| - E_\psi
      |\bvec{p}_b| \cos{\vartheta} \right)^2 \right] + \lambda \left[ -E_b
    |\bvec{k}| + E_\psi |\bvec{p}_b| \cos{\vartheta}\right] \\
  + 1 \left[ |\bvec{p}_b|^2 (1 - \cos^2{\vartheta}) + 2 m_b^2 - 2 E_b
    E_\psi + 2 |\bvec{p}_b| |\bvec{k}| \cos{\vartheta} + \tfrac{1}{2 m_c^2}
    \left(E_b |\bvec{k}| - E_\psi |\bvec{p}_b| \cos{\vartheta} \right)^2 \right]
  \Big\} \\ \times \tfrac{1}{m_c^2} \langle {\cal O}_8^{J/\psi}(^3 P_0) \rangle~.
  \label{ddGamma-3p0-octet}
\end{multline}
In the above, 1 denotes the corresponding terms that contribute to the
spectrum but have no explicit $\lambda$ dependence.
As a consistency check for our results derived so far, 
we choose $|\bvec{p}_b| = 0$, which 
corresponds to the $b$ quark rest frame. Integrating over $\varphi$, 
$\cos{\vartheta}$ and $|\bvec{k}|$, we obtain for the color singlet and octet
contributions to the polarized decay width
\begin{multline}
  \Gamma_1(b \to J/\psi(\lambda) + X) = \frac{1}{96 \pi} G_F^2
  |V_{cb}^{\ast} V_{cf}|^2 \frac{(m_b^2 - 4 m_c^2)^2}{m_b^2 m_c}
  \tfrac{1}{9} C_{[1]}^2 \\
  \times \Big[ \left((m_b^2 - 4 m_c^2) \delta_{\lambda 0}  + 4 m_c^2 (1 - \lambda)
    \right) \langle {\cal O}_1^{J/\psi}(^3 S_1) \rangle \Big]
\end{multline}
and
\begin{multline}
  \Gamma_8(b \to J/\psi(\lambda) + X) = \frac{1}{576 \pi} G_F^2
  |V_{cb}^{\ast} V_{cf}|^2 \frac{(m_b^2 - 4 m_c^2)^2}{m_b^2 m_c}
  C_{[8]}^2 \\
  \times \Big[ \left((m_b^2 - 4 m_c^2) \delta_{\lambda 0} + 4 m_c^2 (1 - \lambda)
    \right) \langle {\cal O}_8^{J/\psi}(^3 S_1) \rangle + m_b^2 \langle
    {\cal O}_8^{J/\psi}(^1 S_0) \rangle \\ 
   + 3 \left( (m_b^2 - 4 m_c^2)
      \delta_{\lambda 0} + m_b^2 + 4 m_c^2 (1 - \lambda) \right)
    \tfrac{1}{m_c^2} \langle {\cal O}_8^{J/\psi}(^3 P_0) \rangle
    \Big]~,
\end{multline}
respectively. This agrees with the result of Fleming et 
al.~\cite{Fleming:1997pt}.
\subsection{Excited quarkonium states}
\label{section-other-quarkonia}
In the previous section, we applied the NRQCD factorization
formalism to $J/\psi$ production in $b$ quark decays, but with
relatively small modifications we can equally well apply our
calculation to other quarkonium states with $J = 1$; as for example $\psi'$ 
and $\chi_{c1}$, which are $2S$ and $1P$ states of charmonium, respectively.

For $\psi'$ production from $b$ quark decays, only the $J/\psi$ 
matrix elements have to be replaced by $\psi'$ matrix elements,
whereas the short-distance coefficients are not affected by this;
\begin{equation}
  \langle {\cal O}_n^{J/\psi}(^{2s+1} L_J) \rangle \rightarrow
  \langle {\cal O}_n^{\psi'}(^{2s+1} L_J) \rangle~.
\end{equation}
Besides, the inclusive decay width to $\chi_{c1}$ requires minor 
modifications, and the formalism is similar to the calculation developed for 
$J/\psi$. 
$\chi_{c1}$, being a $^3 P_1$ state of charmonium, to lowest order
receives contributions from the $\langle {\cal O}_1^{\chi_{c1}} (^{3}
P_1) \rangle$ and $\langle {\cal O}_8^{\chi_{c1}}
(^{3} S_1) \rangle$ matrix elements, other contributions
are down by at least $v^2$. The fact that to lowest order, a color
octet matrix element significantly contributes to the spectrum also explains 
the difficulties to describe $P$ wave quarkonium production in the
framework of quark potential models. However, in the framework of
NRQCD these problems are not completely resolved, even at NLO, since
one faces the task of describing the production of all three
$\chi_{cJ}$ states with one set of matrix elements. This is so,
because the matrix elements 
are related to each other by heavy quark spin symmetry. Usually, they are
expressed in terms of $\chi_{c0}$ matrix elements:
\begin{align}
  \langle {\cal O}_n^{\chi_{cJ}} (^{3} S_1) \rangle =& (2 J + 1)
  \langle {\cal O}_n^{\chi_{c0}} (^{3} S_1) \rangle ~.\\
    \langle {\cal O}_n^{\chi_{cJ}} (^{3} P_J) \rangle =& (2 J + 1)
  \langle {\cal O}_n^{\chi_{c0}} (^{3} P_0) \rangle~.
  \label{equation-chi_c-3pj-relation}
\end{align}
Explicitly, the contributions to the differential decay width for $b
\to \chi_{c1} + X$ are 
\begin{multline}
  E_\chi E_b \frac{d^3 \Gamma_1[^3 P_0]}{d k^3} = \frac{1}{2 \pi^2}
  G_F^2 |V_{cb}^{\ast} V_{cf}|^2  \tfrac{1}{9} C_{[1]}^2 m_c \delta ((p_b - k)^2 - m_f^2) \\
  \Big\{ \delta_{\lambda 0} \left[ \tfrac{1}{2} |\bvec{p}_b|^2 (1 -
    \cos^2{\vartheta}) - \tfrac{1}{2 m_c^2} \left(E_b |\bvec{k}| - E_\chi
      |\bvec{p}_b| \cos{\vartheta} \right)^2 \right] + \lambda \left[ -E_b
    |\bvec{k}| + E_\chi |\bvec{p}_b| \cos{\vartheta}\right] \\
  + 1 \left[ |\bvec{p}_b|^2 (1 - \cos^2{\vartheta}) + 2 m_b^2 - 2 E_b
    E_\chi + 2 |\bvec{p}_b| |\bvec{k}| \cos{\vartheta} + \tfrac{1}{2 m_c^2}
    \left(E_b |\bvec{k}| - E_\chi |\bvec{p}_b| \cos{\vartheta} \right)^2 \right]
  \Big\} \\ \times \tfrac{1}{m_c^2} \langle {\cal O}_1^{\chi_{c0}}(^3 P_0) \rangle~,
  \label{ddGamma-chic1-singlet}
\end{multline}
\begin{multline}
  E_\chi E_b \frac{d^3 \Gamma_8[^3 S_1]}{d k^3} =  \frac{1}{12 \pi^2}
  G_F^2 |V_{cb}^{\ast} V_{cf}|^2  C_{[8]}^2 m_c \delta ((p_b - k)^2 - m_f^2) \\
  \Big\{ \delta_{\lambda 0} \left[ \tfrac{1}{2 m_c^2} \left( E_b |\bvec{k}|
      - E_\chi |\bvec{p}_b| \cos{\vartheta} \right)^2 - |\bvec{p}_b|^2
    (1 - \cos^2{\vartheta}) \right] + \lambda \left[ -E_b |\bvec{k}|
    + E_\chi |\bvec{p}_b| \cos{\vartheta} \right] \\
  + 1 \left[ m_b^2 + |\bvec{p}_b|^2 (1 - \cos^2{\vartheta}) - E_b
    E_\chi + |\bvec{p}_b| |\bvec{k}| \cos{\vartheta} \right] \Big\}
  \times \langle {\cal O}_8^{\chi_{c0}}(^3 S_1) \rangle~.
  \label{ddGamma-chic1-octet}
\end{multline}
\subsection{Feed-down channels: $H \to J/\psi$}
\label{section-feeddown-channels}
Apart from direct $J/\psi$ production which accounts for roughly $70\%$ 
of the $J/\psi$ from $B$ decays, there are 
contributions from feed-down channels. Following a few simplifying 
assumptions as described in \cite{Braaten:1999qk, Kniehl:1999vf}, 
it is possible to incorporate the $J/\psi$ production from these
feed-down channels.

The first step towards this is to calculate the momentum spectra for
the excited charmonium states. In the case of
$\psi'$, because it is an $S$ state the procedure is the same as for $J/\psi$, 
and for the $\chi_{c1}$ production rate, the necessary modifications are very
modest as stated in section \ref{section-other-quarkonia}.
Next, we need to evaluate the production rate of $J/\psi$ from $\psi'$ and
$\chi_{c1}$ decays. It is assumed that in the
excited charmonium decays, the three-momentum is transfered completely to the
$J/\psi$, i.e.,~the $\psi'$ and $\chi_{c1}$ differential production
cross sections are simply multiplied by their experimental branching
fraction to $J/\psi$ final states. The different helicity states are
taken care of by additionally weighting the helicity dependent production
rates for $\psi'_\lambda$ and $\chi_{c1(\lambda)}$ with
probabilities $P(H_\lambda \to J/\psi_{\lambda'})$ where $H = \psi',
\chi_{c1}$, which describe the transition of a $\psi'$ or
$\chi_{c1}$ in helicity state $\lambda$ to a $J/\psi$ in helicity
state $\lambda'$, respectively.

$\psi'$ dominantly decays hadronically into $J/\psi$, and since no spin
flips are observed the polarization is unchanged by this
process \cite{Braaten:1999qk}. Thus we have $P(\psi'_{\lambda} \to
J/\psi_{\lambda'}) = \delta_{\lambda \lambda'}$. For the $\chi_{c1}$ state,
the situation is somewhat different, because it decays radiatively into
$J/\psi$. The transition probabilities have been determined to be
\cite{Cho:1995gb}: $P(\chi_{c1(0)} \to J/\psi_{0}) = 0$,
$P(\chi_{c1(\pm 1)} \to J/\psi_{0}) = \tfrac{1}{2}$, $P(\chi_{c1(0)}
\to J/\psi_{\pm 1}) = 1$ and  $P(\chi_{c1(0)} \to J/\psi_{\pm 1}) =
\tfrac{1}{2}$. Generically, the $J/\psi$ production
through feed-down channels can be summarized as follows
\begin{equation}
  d \Gamma(B \to H_\lambda \to J/\psi_{\lambda'}) = d \Gamma(B \to H_\lambda )
  \cdot Br(H \to J/\psi) \cdot P(H_\lambda \to J/\psi_{\lambda'})~,
\end{equation}
where the inclusive parts $X$ have not been noted in the transitions.
Apart from the above mentioned feed-down channels, there can also 
be radiative transitions from $\chi_{c0}$ and $\chi_{c2}$ which have 
been observed. Their production rates or branching fractions to $J/\psi$ are 
small compared to the ones of $\psi'$ and $\chi_{c1}$ and hence have
been neglected in our analysis.
\section{The hadronic decay $B \to J/\psi + X$}
\label{parton-model}
The results derived in section \ref{b-to-jpsi-x}, 
describe $J/\psi$, or more generally, charmonium production from a free $b$ 
quark decay. In the $b$ rest frame, $J/\psi$ is produced
with fixed momentum, because the kinematic implications of soft
gluon emission in the $J/\psi$ production process are 
neglected. Therefore, at leading order in $\alpha_s$, the momentum 
distribution of the charmonium state results only from the Fermi motion of 
the $b$ quark in the $B$ meson. To incorporate these bound state effects of 
the $B$ meson we adopt the PM as introduced in
\cite{Palmer:1997wv,Jin:1994vc,Jin:1995qc}. An application to semi-leptonic 
$B$ decays was first proposed in \cite{Bareiss:1989my}.

In all our calculations so far, the $b$ quark occurs exclusively in the 
$b \to q_f$ transition, described by the tensor $W_{\mu \nu}$ in
(\ref{eq-w-tensor}). Introducing light-cone dominance as in
\cite{Jin:1994vc, Jin:1995qc}, it is possible to relate the hadronic
$B \to X_f$ transition to the heavy quark parton distribution
function (PDF) $f(x)$. In contrast to semi-leptonic decays, in inclusive
charmonium production the momentum transfer $q^2$ is fixed, if one
neglects the kinematics of soft gluon emission of the final charmonium
state. Due to the on-shell condition for the charmonium, we have
${q^2 = k^2 = M_\psi^2 \approx 10}$~GeV$^2$ which justifies the
light-cone dominance assumption.

At the computational level, the $W_{\mu \nu}$ tensor is
modified in two ways; (i) the quark momentum $p_b$ is replaced by the 
fraction of the $B$ meson momentum $x p_B$, and (ii) the entire partonic 
structure is folded with the PDF $f(x)$ for the heavy $b$ quark.
Thus, we obtain \cite{Palmer:1997wv},
\begin{equation}
  W_{\mu \nu} = 4(S_{\mu \sigma \nu \lambda} - i \epsilon_{\mu \sigma
  \nu \lambda}) \int_0^1 dx f(x) p_B^\lambda(x p_B - k)^\sigma
  \varepsilon[(x p_B - k)_0] \delta[(x p_B - k)^2 - m_f^2]~,
  \label{pm-hadronic-tensor}
\end{equation}
with the sign function
\begin{equation}
  \varepsilon(x) = \left\{\begin{array}{c} +1, \qquad x \ge 0 \\ -1,
  \qquad x < 0 \end{array} \right.~.
\end{equation}
This modification leaves the $T_{\mu \nu}^{(n)}$ tensors unchanged,
they remain as in (\ref{eq-t-tensor-singlet}) and
(\ref{eq-t-tensor-octet}).
The distribution function dependence on the single scaling variable,
$x$, is a consequence of light-cone dominance. In this framework, the
distribution function is obtained as the Fourier transform of the
reduced bi-local matrix element at light-like separations, hence
\begin{equation}
  f(x) = \frac{1}{4 \pi M_B^2} \int d(y\cdot p_B) \left. e^{ix(y\cdot p_B)}
  \langle B | \bar{b}(0) \dirac{p}_B (1 - \gamma_5) b(y) | B \rangle
  \right|_{y^2=0}~,
\end{equation}
as shown in \cite{Jin:1995qc}.

\begin{figure}[htb]
  \begin{center}
    \epsfig{file=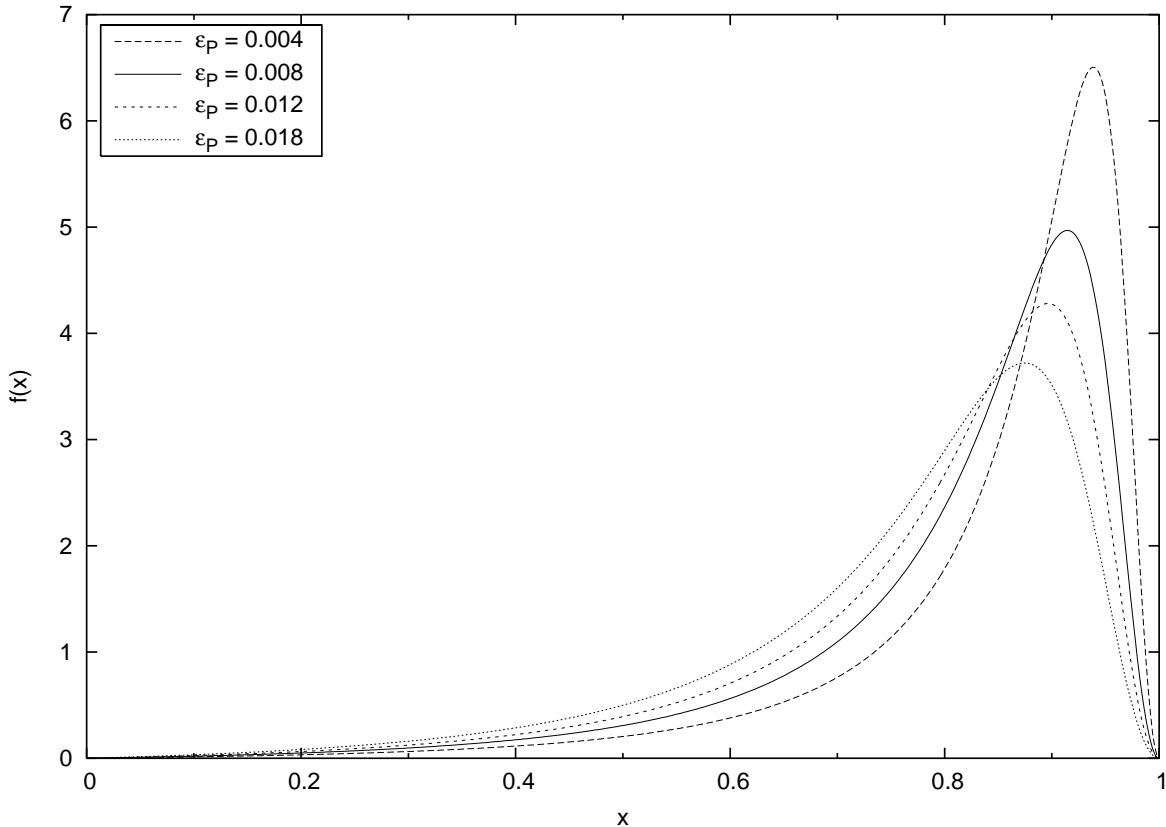,angle=-90,width=16cm}
  \end{center}
  \caption{Peterson function for $\varepsilon_P = 0.004$, $0.008$, $0.012$ and $0.018$.}
  \label{plot-peterson-function}
\end{figure}

In an infinite momentum frame, the distribution function is exactly
the fragmentation function for a high energy $b$ quark to  fragment
into a $B$ meson \cite{Bareiss:1989my}. Hence, the Peterson functional
form \cite{Peterson:1983ak}, can be adopted as a distribution function
for the heavy $b$ quark inside the $B$ meson, with
\begin{equation}
  f(x) = N_\varepsilon \frac{x(1-x)^2}{[(1-x)^2 + \varepsilon_P x]^2}~.
  \label{peterson-functional}
\end{equation}
Here, $f(x)$, is a one parameter function with a free parameter, 
$\varepsilon_P$, while $N_\varepsilon$ is a normalization
factor defined such that,
\begin{equation}
  \int_0^1 dxf(x) = 1~,
\end{equation}
i.e.,~with unit probability there is a $b$ quark in the $B$ meson.
In the parameter range that is usually chosen, $\varepsilon_P \sim
10^{-3} \ldots 10^{-2}$, $f(x)$ peaks at large values of $x$; a behavior that
has been determined elsewhere 
\cite{Bjorken:1978md,Brodsky:1981se}. For completeness, in 
figure \ref{plot-peterson-function}, the $x$ dependence of $f(x)$ is shown 
for four different values of $\varepsilon_P$.

As a useful consequence of the PM, the $b$ quark mass, $m_b$, is
replaced by the $B$ meson mass $M_B$ removing the uncertainty in the
quark mass. On the other hand, the parameter $\varepsilon_P$ now
carries quite a large uncertainty, and in some form, we have traded one 
uncertainty for another. For comparison with other calculations,
it can be useful to define an effective $b$ quark mass \cite{Lee:1996rq}
\begin{equation}
  m_b = \langle x \rangle M_B~,
  \label{eq-effective-b-mass}
\end{equation}
where
\begin{equation}
  \langle x \rangle = \int_0^1 dx~x f(x)~.
\end{equation}
To illustrate the dependence of 
$\langle x \rangle$ and $m_b$ on $\varepsilon_P$, we give their values for 
different choices of $\varepsilon_P$ in table \ref{table-effective-b-mass} 
setting $M_B = 5.279$~GeV.
\begin{table}[htb]
  \begin{center}
    \begin{tabular}{|c|c|c|}
      \hline 
      $\varepsilon_P$ & $\langle x \rangle$ & $m_b$ in GeV\\
      \hline
      $0.004$ & $0.85$ & $4.48$ \\
      $0.008$ & $0.81$ & $4.29$ \\
      $0.012$ & $0.78$ & $4.16$ \\
      $0.018$ & $0.76$ & $4.02$ \\
     \hline
    \end{tabular}
  \end{center}
  \caption{Effective $b$ quark mass $m_b$ and expectation value
      $\langle x \rangle$ of $f(x)$ for various $\varepsilon_P$ and $M_B =
      5.279$~GeV.}
  \label{table-effective-b-mass}
\end{table}

As expected, the results of (\ref{ddGamma-3s1-singlet} --
\ref{ddGamma-3p0-octet}) get modified due to the new hadronic
tensor $W_{\mu \nu}$ in (\ref{pm-hadronic-tensor}). The integration
over the scaling variable $x$ is straight
forward and therefore can be performed immediately. The generic
integral $I$ under consideration is of the form
\begin{equation}
  I = \int_0^1 dx~G(x) \theta[(x p_B - k)_0] \delta[(x p_B - k)^2 -
  m_f^2]~, \label{x-integral}
\end{equation}
where $G(x)$ is an arbitrary function of the integration variable
$x$. Note the presence of the $\theta$ function which ensures that the final state
$q_f$ quark has positive energy\footnote{The $\theta$ function thus
  effectively replaces the $\varepsilon$ function in
  (\ref{pm-hadronic-tensor}).}. This is required, because in the 
inclusive approach $q_f$ has to hadronize, finally giving a hadronic
state $X_f$ which of course has to have positive energy.

To solve the integral, we introduce a new integration variable
according to
\begin{equation}
  (x p_B - k)^2 - m_f^2 = M_B^2\left[z^2 - \frac{(p_B k)^2}{M_B^4} +
    \frac{M_\psi^2 - m_f^2}{M_B^2} \right]
\end{equation}
with
\begin{equation}
  z(x) = x - \frac{p_B k}{M_B^2}~.
\end{equation}
This translates (\ref{x-integral}) into an integral over $z$ and 
\begin{equation}
  I = \int_{z(0)}^{z(1)} dz G(z + z_0) \theta[(z + z_0) p_{B(0)} - k_0]
  \frac{1}{2 M_B^2 z_1} \left( \delta[z - z_1] + \delta[z +
    z_1]\right)~, \label{y-integral}
\end{equation}
with
\begin{align}
  z_0 &= p_B k / M_B^2~, \\
  z_1 &= \sqrt{(p_B k)^2 - M_B^2 (M_\psi^2 - m_f^2)}/ M_B^2~.
\end{align}
Here, we have used the standard identities for the $\delta$
distribution. The second $\delta$ function in (\ref{y-integral})
does not contribute due to the $\theta$ function which restricts
contributions of the integral to the argument of the first $\delta$
function. Therefore, (\ref{x-integral}) can finally be rewritten as
\begin{equation}
  I = \frac{1}{2 M_B^2 z_1} G(x_{+}) \theta(x_{+}) \theta(1 - x_{+})~,
\end{equation}
where we have defined
\begin{equation}
x_{+} = z_0 + z_1 = \frac{1}{M_B^2} \left( p_B k + \sqrt{(p_B k)^2
- M_B^2 (M_\psi^2 - m_f^2)} \right) .
\end{equation}
The two $\theta$ functions express the fact that the scaling variable
$x_+$, since it can be interpreted as the $b$ quark's momentum
fraction within the $B$ meson, is only allowed to vary between 0 and
1.

Inserting the above expressions into (\ref{ddGamma-3s1-singlet} --
\ref{ddGamma-3p0-octet}) and together with the tensor
(\ref{pm-hadronic-tensor}), we obtain the following expressions for
the decay width, sorted by their production mechanism.
\begin{multline}
  E_\psi E_B \frac{d^3 \Gamma_1[^3 S_1]}{d^3 k} =  \frac{1}{12 \pi^2}
  G_F^2 |V_{cb}^{\ast} V_{cf}|^2  \tfrac{1}{9} C_{[1]}^2
  \frac{m_c}{M_B^2} \frac{1}{z_1} f(x_{+}) \theta(x_{+}) \theta(1 -
  x_{+}) \\
  \Big\{ \delta_{\lambda 0} x_+ \left[ \tfrac{1}{2 m_c^2} \left( E_B |\bvec{k}|
      - E_\psi |\bvec{p}_B| \cos{\vartheta} \right)^2 - |\bvec{p}_B|^2
    (1 - \cos^2{\vartheta}) \right] + \lambda \left[ -E_B |\bvec{k}|
    + E_\psi |\bvec{p}_B| \cos{\vartheta} \right] \\
  + 1 \left[ x_+ M_B^2 + x_+ |\bvec{p}_B|^2 (1 - \cos^2{\vartheta}) -
  E_B E_\psi + |\bvec{p}_B| |\bvec{k}|
    \cos{\vartheta} \right] \Big\} \times \langle {\cal
  O}_1^{J/\psi}(^3 S_1) \rangle~.
  \label{ddGamma-3s1-singlet-pm}
\end{multline}
\begin{multline}
  E_\psi E_B \frac{d ^3\Gamma_8[^1 S_0]}{dk^3} =  \frac{1}{72 \pi^2}
  G_F^2 |V_{cb}^{\ast} V_{cf}|^2  C_{[8]}^2 \frac{m_c}{M_B^2}
  \frac{1}{z_1} f(x_{+}) \theta(x_{+}) \theta(1 - x_{+}) \\
   \Big\{ 1 \left[ -E_B E_\psi + |\bvec{p}_B| |\bvec{k}| \cos{\vartheta} -
    x M_B^2 + \tfrac{1}{2 m_c^2} x_{+} \left( E_B E_\psi - |\bvec{p}_B| |\bvec{k}|
      \cos{\vartheta} \right)^2 \right] \Big\} \times \langle {\cal
  O}_8^{J/\psi}(^1 S_0) \rangle~.
  \label{ddGamma-1s0-octet-pm}
\end{multline}
\begin{multline}
  E_\psi E_B \frac{d ^3\Gamma_8[^3 S_1]}{dk^3} =  \frac{1}{72 \pi^2}
  G_F^2 |V_{cb}^{\ast} V_{cf}|^2  C_{[8]}^2 \frac{m_c}{M_B^2}
   \frac{1}{z_1} f(x_{+}) \theta(x_{+}) \theta(1 - x_{+}) \\
  \Big\{ \delta_{\lambda 0} x_+ \left[ \tfrac{1}{2 m_c^2} \left( E_B |\bvec{k}|
      - E_\psi |\bvec{p}_B| \cos{\vartheta} \right)^2 - |\bvec{p}_B|^2
    (1 - \cos^2{\vartheta}) \right] + \lambda \left[ -E_B |\bvec{k}|
    + E_\psi |\bvec{p}_B| \cos{\vartheta} \right] \\
  + 1 \left[ x_+ M_B^2 + x_+ |\bvec{p}_B|^2 (1 - \cos^2{\vartheta}) -
  E_B E_\psi + |\bvec{p}_B| |\bvec{k}| \cos{\vartheta} \right] \Big\}
  \times \langle {\cal O}_8^{J/\psi}(^3 S_1) \rangle ~.
  \label{ddGamma-3s1-octet-pm}
\end{multline}
\begin{multline}
  E_\psi E_B \frac{d ^3\Gamma_8[^3 P_0]}{dk^3} = \frac{1}{24 \pi^2}
  G_F^2 |V_{cb}^{\ast} V_{cf}|^2  C_{[8]}^2 \frac{m_c}{M_B^2}
  \frac{1}{z_1} f(x_{+}) \theta(x_{+}) \theta(1 - x_{+}) \\
  \Big\{ \delta_{\lambda 0} \left[ \tfrac{1}{2} |\bvec{p}_B|^2 (1 -
    \cos^2{\vartheta}) - \tfrac{1}{2 m_c^2} x_+ \left(E_B |\bvec{k}| - E_\psi
      |\bvec{p}_B| \cos{\vartheta} \right)^2 \right] + \lambda \left[
    -E_B |\bvec{k}| + E_\psi |\bvec{p}_B| \cos{\vartheta}\right] \\
  + 1 \left[ \vphantom{\tfrac{1}{m_c^2}} x_+ |\bvec{p}_B|^2 (1 -
    \cos^2{\vartheta}) + 2 x_+ M_B^2 - 2 E_B E_\psi + 2 |\bvec{p}_B|
    |\bvec{k}| \cos{\vartheta} \right. \\ 
  \left. + \tfrac{1}{2m_c^2} x_+ \left(E_B |\bvec{k}| - E_\psi |\bvec{p}_B|
      \cos{\vartheta} \right)^2 \right]
  \Big\} \times \tfrac{1}{m_c^2} \langle {\cal O}_8^{J/\psi}(^3 P_0) 
  \rangle~.
  \label{ddGamma-3p0-octet-pm}
\end{multline}
The distribution function $f(x)$, being defined in an infinite
momentum frame, as well as the incoherence assumption restrict the
PM application to $B$ mesons with large, if not infinite,
three-momenta. To circumvent this problem, in the case of unpolarized decay,
a Lorentz invariant quantity, $E\cdot d\Gamma$, is usually
constructed and subsequently evaluated in an arbitrary reference
frame. This strategy was adopted in the calculation of semi-leptonic
$B$ decays (e.g.~\cite{Bareiss:1989my,Jin:1994vc}) and also in
\cite{Palmer:1997wv} where the PM was first applied to inclusive
hadronic $B$ decays. This enabled the authors to employ their calculations
to $B$ decays at the CLEO experiment where $B$ mesons are produced
almost at rest. However, this is not possible in the case of polarized
production cross sections, since they are frame dependent. Therefore one has
to go to a large momentum frame to fulfill the requirements of the
PM. This brings us to the description and application of our
results to the Tevatron.

\section{$B \to J/\psi + X$ at the Tevatron}
\label{section-tevatron}
At the Tevatron, $B$ mesons are not produced with fixed momentum as in 
$\Upsilon$ decays (e.g.,~at CLEO), but in fragmentation
mode. Therefore, one has to deal with a momentum distribution
of the $B$ mesons, expressed by the $p_T$ and $y_B$ ($B$ rapidity) dependent 
double differential $B$ production cross section. The double
differential $B$ cross section ${d^2\sigma(p\bar{p} \to B + X)/(dp_T
  dy_B)}$ has to be folded with the normalized semi-inclusive
differential decay spectrum, ${1/\Gamma_B \cdot d^3 \Gamma (B
  \to J/\psi + X')/d k^3}$, to obtain the desired differential
production cross section for $J/\psi$.

The CDF collaboration at Tevatron has already measured $J/\psi$ and $\psi'$ 
production from $B$ decays \cite{Daniels:1997,Abe:1997jz} as well as their 
polarization \cite{Affolder:2000nn}. For the latter we are not aware of any
theoretical predictions that take into account bound state effects.
Our results presented here (with some kinematic modifications) are directly 
applicable to the Tevatron experimental setup. We proceed to discuss the 
required kinematics to suit the Tevatron and make a comparison of our
predictions with the available data.

So far the absolute
values of the three-momenta $|\bvec{p}_B|$, $|\bvec{k}|$ and the 
polar angle $\vartheta$ between $\bvec{p}_B$ and $\bvec{k}$ have
formed the set of kinematic variables in our calculations. For these,
we will trade with a different set: the transverse momenta $p_T$ and
$k_T$, the rapidities $y_B$ and $y$ and the azimuthal angle
$\phi$. $p_T$ and $y_B$ belong to the $B$ meson, whereas $k_T$, $y$
and $\phi$ describe the kinematics of $J/\psi$. 
Following this, the relations between
the old and new variables are given to be 
\begin{align}
  |\bvec{k}|  =& \left(k_T^2 + \sinh^2{y} (k_T^2 +
    M_\psi^2)\right)^{1/2} ~,\nonumber\\
  |\bvec{p}_B| =&  \left(p_T^2 + \sinh^2{y_B} (p_T^2 +
    M_B^2)\right)^{1/2} ~,\\
  \cos{\vartheta} =& \frac{1}{|\bvec{k}| |\bvec{p}_B|}
  \left(\sinh{y} \sqrt{k_T^2 + M_\psi^2} \cdot \sinh{y_B}
    \sqrt{p_T^2 + M_B^2} - k_T p_T \cos{\phi} \right) ~.\nonumber
\end{align}
Here, the rapidities are defined to be
\begin{equation}
  y = \tfrac{1}{2} \ln{\left(\frac{E + k_{\parallel}}{E -
  k_{\parallel}}\right)} \qquad \textrm{and} \qquad y_B = \tfrac{1}{2}
  \ln{\left(\frac{E_B + p_{\parallel}}{E_B - p_{\parallel}}\right)}~,
\end{equation}
with $k_\parallel$ and $p_\parallel$ being the momentum components
parallel to the beam line. The Jacobian determinant for the coordinate
transformation of the $\psi$ variables is
\begin{equation}
\left|\frac{\partial(k_x, k_y, k_z)}{\partial(k_T, y, \phi)}
\right| = k_T \sqrt{k_T^2 + M_\psi^2} \cosh{y}~.
\end{equation}
\subsection{Cross section and the $\alpha$ parameter}
\label{bcross}
In order to make a comparison with data, we have to take in to account the 
relevant experimental constraints and cuts. In the case of 
$J/\psi$ and $\psi'$ production, the cross section is rapidity integrated 
over the region $|y| \le 0.6$ \cite{Daniels:1997,Abe:1997jz,Affolder:2000nn}. 
For direct $J/\psi$ production the cross section is expressed as
\begin{equation}
  \frac{d \sigma (J/\psi)}{d k_T} = \int d p_T
  \int d y_B \frac{d^2 \sigma (B)}{d p_T d y_B} \int d \phi
  \int d y \left|\frac{\partial(k_x, k_y, k_z)}{\partial(k_T, y, \phi)}
  \right| \frac{1}{\Gamma_B} \frac{d^3 \Gamma}{d k^3}(p_T, y_B, k_T, y, \phi)~.
  \label{sigma-convolution}
\end{equation}
As already discussed in section \ref{section-feeddown-channels}, apart
from the direct production also the feed-down channels
contribute to $J/\psi$ production. These can be incorporated 
under the assumptions made in section \ref{section-feeddown-channels}
and hence the decay rates to $J/\psi$ just have to be summed for the
direct and the feed-down channels to obtain the final result.
Using (\ref{sigma-convolution}), we can evaluate the usual polarization 
parameter, $\alpha$, which is experimentally accessible
with the help of a fit to the angular distribution in the di-lepton decays
of $J/\psi$. The angular differential decay
spectrum has the following form
\begin{equation}
  \frac{d \Gamma(\psi \to l^+ l^-)}{d \cos{\theta}} \propto 1 + \alpha
  \cos^2{\theta}~,
  \label{alpha-def} 
\end{equation}
where the angle $\theta$ is defined in the $\psi$ rest frame in
which the $z$ axis is aligned with the direction of motion of the
$J/\psi$ in the $B$ rest frame.

Theoretically, $\alpha$ is expressed
as the ratio of linear combinations of the helicity production rates
for $J/\psi$. Given the decay of a longitudinally polarized vector
particle (helicity $\lambda = 0$) and for a transversely polarized 
state (helicity $\lambda = \pm$), using (\ref{alpha-def}) one immediately 
obtains
\begin{equation}
  \alpha = \frac{\sigma_+ + \sigma_- - 2 \sigma_0}{\sigma_+ + \sigma_-
  + 2 \sigma_0}~.
\end{equation}
The helicity production cross sections $\sigma_\lambda$ are the ones obtained
from (\ref{sigma-convolution}), which makes $\alpha$ a function of $k_T$.  

\subsection{Present status}
With respect to the Tevatron, there are two ways of employing the 
differential $B$ production cross section in our calculation; 
either theoretically or experimentally. 
One approach, is to calculate the cross section by means of 
the QCD improved PM. In this case, the partonic $b$ quark production 
cross sections have to be calculated in perturbative QCD and
subsequently have to be folded with (i) the non-perturbative PDFs of the 
incoming hadrons and (ii) the fragmentation function, describing the 
hadronization process of the $b$ quark. In the second approach, one can 
directly use the measured $B$ production cross section at the Tevatron  
c.m.s.~energy $\sqrt{s} =1.8$~TeV. 

It is known for quite a number of years that the theoretical
description of the $b$ production process fails to reproduce the 
experimental data (see e.g.~\cite{Albajar:1987iu}). The shape of the spectrum 
comes out as desired, but the normalization usually falls short by a factor 
of 2 or more (e.g.~\cite{Albajar:1987iu,Abe:1993sj,Abbott:1999se}). An 
agreement with the data can be achieved if the involved parameters like
the factorization scale, $\mu$, $\Lambda_{\textrm{QCD}}$ and $m_b$ are driven
to rather extreme values. Recently, there has
been a suggestion that the fragmentation function ansatz which has been used
most frequently might not be appropriate \cite{Cacciari:2002pa}.
However, in a next-to-leading order calculation of $B$ meson production in $p
\bar{p}$ collisions, it was shown that it is possible to accommodate the data 
within experimental error bars without fine tuning of the relevant
parameters \cite{Binnewies:1998vm}. In this case, the $b$ fragmentation
function has been fitted to LEP data and the differences in the scales between
LEP and CDF data is accounted for by the usual evolution equations.

At this stage, it is not clear if the corrections due to the $b$ fragmentation
are responsible for the disagreements between theoretical predictions and
experimental data. It may also be that the theory for the $b$ production
mechanism is incomplete. There has been an attempt to account for the
discrepancy involving  physics beyond the standard 
model \cite{Berger:2000mp}.\\

In the present analysis, we adopt the second strategy, i.e., to use an 
experimental fit. We are motivated to this choice, because,
the main concern of this work is not the dynamical mechanism of $B$
meson production in $p \bar{p}$ collisions, but their subsequent decay
into quarkonium states. A strong argument for this being, the
available data have been extracted using exclusive $B^{\pm} \to J/\psi
+ K^{\pm}$ decays with the CDF detector at the Tevatron
\cite{Abe:1995dv,Acosta:2001rz}. The same group has also measured the
$J/\psi$ and $\psi'$ production cross section \cite{Abe:1997jz} and
their polarization \cite{Affolder:2000nn}. Therefore systematic errors
in the analysis should be of less significance and the comparison
between theory and experiment will be more meaningful. The problems due to 
the lack of experimental data for the double differential $B$ production 
cross section will be discussed later in this section. To this end, we
first derive a useful algebraic fit to the production cross section in the
following subsection.
\subsection{A two parameter fit}
It has been shown that the
differential production cross section with respect to the $B$
transverse momentum, $p_T$, exhibits a simple power law behavior
\cite{Nason:1999ta} with a power in the range of $-3$ and $-5$. We
therefore choose a power law ansatz of the type
\begin{equation}
  \frac{d \sigma}{d p_T} = A \cdot p_T^n~,
\end{equation}
with $A$ and $n$ being the two fit parameters. Here, we minimize
the relative rather than the absolute deviation square. This is done
to make the functional form for the fit to
correctly reproduce the cross section for both low and high $p_T$. The
reason being, at high $p_T$, the absolute value of the cross section becomes rather 
small because of the power law behavior.

\begin{figure}[htb]
  \begin{center}
    \epsfig{file=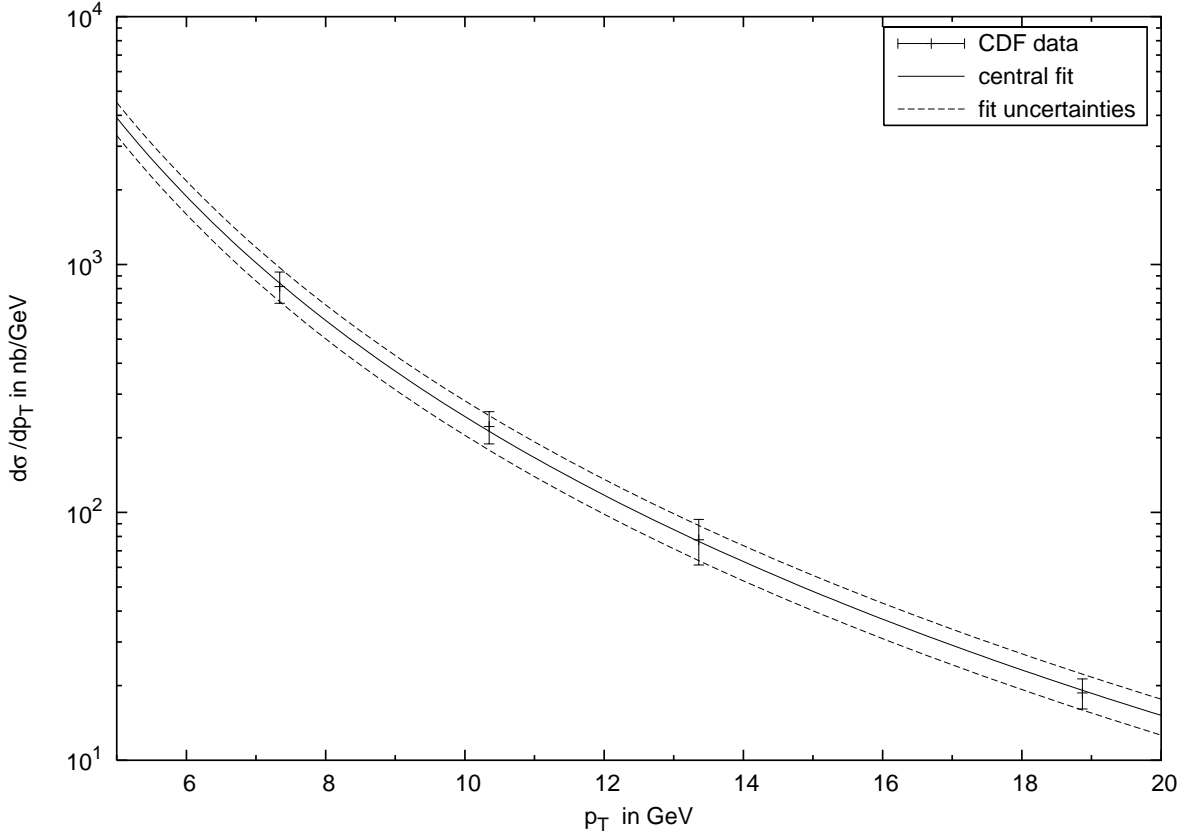,angle=-90,width=16cm}
  \end{center}
  \caption{Experimental data \cite{Acosta:2001rz} and fit results for
    differential $B^+$ production cross section $d\sigma/dp_T$.}
  \label{plot-fit-B-production}
\end{figure}

To estimate the uncertainty due to the experimental $B$ production
cross section, we apply the fit procedure not only to the central
values, but also to the cross section $\pm$~statistical errors, i.e.,~we
fit the ``upper/lower ends of the error bars''. In table
\ref{table-fit-B-production}, the results of the three fits including
their $\chi^2$ values are given, and these are also displayed in figure
\ref{plot-fit-B-production}.
\begin{table}[htb]
  \begin{center}
    \begin{tabular}{|l|c|c|c|}
      \hline 
      & central & upper & lower \\
      \hline
      $A$ & $2.4635 \cdot 10^6$ & $2.8084 \cdot 10^6$ & $2.1363 \cdot 10^6$
      \\
      $n$ & $-4.0049$ & $-3.9984$ & $-4.0187$ \\
      $\chi^2_{\textrm{dof}}$ & $0.166/2$ & $0.437/2$ & $0.125/2$ \\
      \hline
    \end{tabular}
  \end{center}
  \caption{Fit results of the model $A \cdot p_T^n$ to the differential $B^+$
    production cross section $d\sigma/dp_T$ for central, upper and
      lower experimental values.}
  \label{table-fit-B-production}
\end{table}

However, given the kinematics
relevant to the Tevatron, we note that the fit for the experimental 
cross section as an input distribution is only available in 
rapidity integrated form. On the other hand, one needs to convolute the double
differential $B$ production cross section with the normalized
differential decay spectrum for $B \to J/\psi + X$ with respect to
transversal momentum $p_T$, as well as rapidity $y_B$ (see equation
(\ref{sigma-convolution})). In the following, we assume the double 
differential cross section to be only weakly dependent on $y_B$ in the
dominantly contributing $y_B$ range. Later we see that this assumption gets
justified for our setup. The error in the
numerical result of the differential $J/\psi$ production cross section
due to this treatment can be estimated by including a purely phenomenological
$y_B$ dependence of the $B$ production cross section, based on
theoretical calculations. We have tried to reproduce the rapidity
dependence of the double differential cross section $d^2 \sigma (p
\bar{p} \to B + X)/(d p_T dy_B)$ given in 
\cite{Binnewies:1998vm} for small values of $y_B$ and have chosen the
normalization such that the central experimental values of $d \sigma/d
p_T$ in \cite{Acosta:2001rz} were recovered when
integrating over $y_B$ in the range $-1$ to $+1$. In a wide $p_T$ range
($10 \ldots 20$~GeV), we note that, 
$d^2 \sigma/(d p_T dy_B)$ reduces by roughly
$15\%$ between $|y_B| = 0$ and $|y_B| = 1$ and so the factorization
assumption made here seems to be reasonable. A particularly simple 
form for the double differential cross section can be taken to be of the form
\begin{equation}
  \frac{d^2 \sigma}{d p_T dy_B} = (B - C y_B^m) \cdot p_T^n~.
\label{ddc}
\end{equation}
Note that the function in (\ref{ddc}) has to be even in $y_B$. To 
satisfy this, the simplest choice is $m = 2$, which, for not too large 
values of $y_B$, reproduces the shape of the curve in \cite{Binnewies:1998vm} 
sufficiently well. For this choice, we have a few constraints which are to 
be satisfied; $B$ and $C$ have to be chosen such that the 
$15\%$ reduction between $|y_B| = 0$ and $1$ as mentioned here is
accounted for. Furthermore, 
the parameter $A$ has to be recovered when integrating over the $y_B$ interval
$[-1,+1]$. These conditions together fix the parameters unambiguously to be
\begin{align}
  B &= \left(\frac{1}{2} + \frac{1}{x_y} - \frac{1}{m +
  1}\right)^{-1} \left(\frac{1}{2} + \frac{1}{x_y}\right)
  \frac{A}{2} ~, \\
  C &= \left(\frac{1}{2} + \frac{1}{x_y} - \frac{1}{m +
  1}\right)^{-1} \frac{A}{2} ~,
\label{fit}
\end{align}
where $x_y \approx 15 \%$. We use this result to estimate the 
error in our fits for the production cross section when we assume
$y_B$ independence.

\subsection{Cumulative cross sections from all $b$ Hadrons}
The $B$ cross section that we applied in the last section refers only
to $B^+$ production, whereas a $B \to \psi + X$ decay could involve any $B$
meson type and even $\Lambda_b$ baryons. Analysis of $b$ quark
fragmentation fractions are available from LEP data
\cite{Groom:2000in} and from the CDF collaboration (Tevatron)
\cite{Abe:1999ta, Affolder:1999iq} and we use them to include
contributions from $B^0$, $B_s^0$ and $\Lambda_b$ baryons. As a passing remark,
we note that the number of produced $B_c$ mesons is too small to give
a sizeable contribution \cite{Braaten:1993jn}. We make the simplifying
assumption that the shape of
the production cross section is the same for all $b$ flavored hadrons and
hence we multiply the $B^+$ differential cross section by a factor
$F_B$ to include the contributions from the other $b$ hadron
types. This is certainly a good approximation for $B^0$, but ignores the mass
differences for $B^0_s$ ($\Delta_M \sim 100$~MeV) and
$\bar{\Lambda}_b$ ($\Delta_M \sim 340$~MeV). Finally, the net production rate
has to be multiplied by $2$, because at the quark level both $b$  
as well as the charge conjugated decays can give rise to charmonium states.   

Apart from the $B_s^0$ fragmentation fraction $Br(b \to \bar{B}^0_s)$,
which at the Tevatron is measured to be $2\sigma$ above
the LEP result, the extracted values are
compatible within one $1\sigma$. We choose to use the 
Tevatron data \cite{Affolder:1999iq} as central values for our analysis, due to
the possible differences of $b$ fragmentation in
$e^+ e^-$ and $p \bar{p}$ collisions. The experimental data as well as
the combined scaling factor, $F_B$, that we finally use to incorporate the
other $b$ hadron contributions are summarized in table \ref{table-b-fractions}.
\begin{table}[htb]
  \begin{center}
    \begin{tabular}{|l|c|c|}
      \hline 
      fragmentation fraction & CDF notation & CDF value \\
      \hline
      $Br(\bar{b} \to B^+)$ & $f_u$ & $0.375 \pm 0.015$ \\
      $Br(\bar{b} \to B^0)$ & $f_d$ & $0.375 \pm 0.015$ \\
      $Br(\bar{b} \to B^0_s)$ & $f_s$ & $0.160 \pm 0.025$ \\
      $Br(\bar{b} \to \bar{\Lambda}_b)$ & $f_{\textrm{baryon}}$ & $0.090 \pm 0.028$ \\
      \hline
      \multicolumn{2}{|l|}{$\tfrac{1}{2} F_B$} & $2.67 \pm 0.11$\\
      \hline
    \end{tabular}
  \end{center}
  \caption{Experimental results on $b$ fragmentation fractions at the
      Tevatron \cite{Affolder:1999iq} and the combined scaling factor $F_B$.}
  \label{table-b-fractions}
\end{table}
\section{NRQCD matrix elements}
\label{section-nrqcd-matrix-elements}
The most crucial input parameters in the entire calculation are the NRQCD
matrix elements, because their values influence the theoretical predictions
significantly. Numerical values of these matrix elements for $J/\psi$,
$\psi'$ and $\chi_{cJ}$ production that have been published in the
literature are summarized in the appendix (see section \ref{appendix-nrqcd}). 
They have been determined mostly in unpolarized prompt $\psi$ production at
hadron colliders, but also at $ep$ colliders and in fixed target
experiments. Apart from the $^3 P_J$ element, all the other matrix elements 
considered here have to be positive \cite{Fleming:1997pt}. In the following,
we quickly recapitulate the extraction of the color singlet and octet 
matrix elements from various phenomenological data which we shall use.

\subsection{Color singlet matrix elements}
Leading color singlet matrix elements can be obtained in
various ways. A popular approach is to calculate the quarkonium wave
function within a quark potential model, adopting the QCD inspired 
Buchm\"uller-Tye potential \cite{Buchmuller:1981su}.
Alternatively, experimental decay rates of quarkonium states can be
used to obtain numerical values for the color singlet matrix
elements. For $S$ states, the partial decay width, 
$\Gamma(\psi \to e^+ e^-)$, is expressible in terms of the
color singlet matrix elements up to higher corrections in the $v^2$
expansion \cite{Bodwin:1995jh}; and hence,  
$\Gamma(\psi \to e^+ e^-) \propto \langle {\cal O}_1^{\psi}
 (^{3} S_1) \rangle$. Using experimental data from
\cite{Groom:2000in}, we find for $J/\psi$ the values
\begin{equation}
  \langle{\cal O}_1^{J/\psi} (^{3} S_1) \rangle_{(0)} = (0.77 \pm
  0.06)~\textrm{GeV}^3 
  \quad \textrm{and} \quad \langle{\cal O}_1^{J/\psi} (^{3} S_1) \rangle_{(1)} =
  (1.37 \pm 0.10)~\textrm{GeV}^3 ~,
\end{equation}
where the results are displayed without (subscript $(0)$) and with QCD 
corrections (subscript $(1)$),
respectively. The error corresponds only to uncertainties in data. Here, we 
have chosen $\alpha_s(2 m_c) = 0.26$ and
$M_{J/\psi} = 3.097$~GeV. Evaluating the same formula for the $\psi'$,
we obtain
\begin{equation}
  \langle{\cal O}_1^{\psi'} (^{3} S_1) \rangle_{(0)} = (0.44 \pm 0.04)~\textrm{GeV}^3
  \quad \textrm{and} \quad \langle{\cal O}_1^{\psi'} (^{3} S_1) \rangle_{(1)} =
  (0.78 \pm 0.07)~\textrm{GeV}^3~, 
\end{equation}
where $M_{\psi'} = 3.686$~GeV has been adopted. For $M_{\psi'} = 2 m_c
= 3$~GeV, these values reduce by roughly one third.

For $P$ wave charmonium states, the electromagnetic decay $\chi_{cJ}
\to \gamma \gamma$ for $J = 0$ and $J = 2$ are suitable to fix the
leading color singlet matrix elements $\langle {\cal
  O}_1^{\chi_{cJ}} (^{3} P_J) \rangle$, which are related to each other
through (\ref{equation-chi_c-3pj-relation}). For the
$\chi_{c0}$ decay, the precision of the available data is lower than for 
the $\chi_{c2}$ decay \cite{Groom:2000in}. Therefore, the latter one is 
adopted more often to fix the corresponding 
$\langle {\cal O}_1^{\chi_{c0}} (^{3} P_0)
\rangle$ matrix element. Including QCD corrections to lowest order,
the decay width can then be expressed in terms of the leading 
color singlet matrix element \cite{Bodwin:1995jh}. In this case, we
have, 
$\Gamma(\chi_{c2} \to \gamma \gamma) \propto \langle {\cal O}_1^{\chi_{c0}}
 (^{3} P_0) \rangle$. Numerically, we find 
\begin{equation}
  \langle{\cal O}_1^{\chi_{c0}} (^{3} P_0) \rangle_{(0)} = (3.9 \pm
  1.4) \cdot 10^{-2}~\textrm{GeV}^5 \textrm{ and } \langle{\cal O}_1^{\chi_{c0}} (^{3}
  P_0) \rangle_{(1)} = (7.0 \pm 2.5) \cdot 10^{-2}~\textrm{GeV}^5
\end{equation}
for the tree level and the QCD improved calculation,
respectively. Here $M_{\chi_{c2}} = 3.556$~GeV was chosen,
$M_{\chi_{c2}} = 2 m_c = 3$~GeV results in a $50\%$ reduction of
these values.

Alternatively, the decays of $\chi_{c0}$ and $\chi_{c2}$ into light
hadrons have been used to fix the above matrix elements \cite{Bodwin:1992ye}
with results that are compatible with the ones obtained above.

A third possibility which also is adopted quite often, is to calculate
the matrix elements on the lattice \cite{Bodwin:1996tg}. In principle
all the three methods give matrix element of comparable size, in particular, 
if one chooses the mean value of the leading order and the QCD improved 
result in the experimental analysis.

\subsection{Color octet elements}
In the case of $J/\psi$ and $\psi'$ prompt production data from fixed
target as well as collider experiments have been used for extracting the
color octet elements. Due to the different $k_T$ dependence of the matrix 
elements contributions, it is possible to fit their values to the unpolarized
differential prompt $\psi$ production cross section, $d \sigma/d
k_T$. At large $k_T$, due to gluon fragmentation dominance, as described
in the beginning, the $^3 S_1$ matrix elements become the dominant source 
for $\psi$ production. On the other hand, the gluon fusion process gives 
rise to contributions that are proportional to the $^1 S_0$ and the $^3 P_0$
color octet matrix elements. Since these two contributions
exhibit a very similar $k_T$ dependence, the matrix elements cannot be fixed
individually in the fit procedure. Instead, it was suggested to fix
a linear combination of both \cite{Cho:1996vh,Cho:1996ce}, which is
conventionally denoted by 
\begin{equation}
  M_r^\psi = \langle {\cal O}_8^{\psi} (^{1} S_0) \rangle + \frac{r}{m_c^2}
  \langle {\cal O}_8^{\psi} (^{3} P_0) \rangle~,
  \label{Mr-definition}
\end{equation}
where $r$ is empirically determined and lies in the range $3
\ldots 3.5$ for hadroproduction.

The fitting procedure for the NRQCD cross section for 
hadroproduction, as well as for electro- and
photoproduction involves PDFs as theoretical input of the
calculation. Hence, a theoretical uncertainty is introduced, because
for different PDF sets the values of the matrix elements differ
quite severely (see e.g.~\cite{Leibovich:2000qi}).

There have also been efforts to individually fix the $^1 S_0$ and $^3
P_0$ color octet matrix elements for $\psi$ production by fitting
their values to the $\psi$ momentum spectrum from $B$ decays
\cite{Palmer:1997wv,Kniehl:1999vf} for the spectra 
as measured by the CLEO collaboration \cite{Balest:1995jf}. However, 
at the high momentum end of the spectrum, as was noted 
in \cite{Kniehl:1999vf}, one encounters difficulties because of effects of 
resonant two- and three-particle decay channels in the spectrum. In the
context of the Tevatron, since only the absolute normalization
of the spectrum is of primary importance, it is best to fit for the
branching fraction. This procedure was first used in 
\cite{Kniehl:1999vf} and the matrix elements were fixed to the
inclusive branching fractions, $Br(B \to \psi + X)$ for the CLEO data. In our
analysis, we adopt this procedure as mentioned in the following subsection.

\subsection{Prescription for analysis}
\label{section-analysis-strategy}
All the difficulties mentioned above, make precise theoretical predictions
involving the matrix elements to be very hard. A possibility to constrain  
especially, the $\langle
{\cal O}_8^{\psi} (^{1} S_0) \rangle$ and the $\langle {\cal
  O}_8^{\psi} (^{3} P_0) \rangle$ matrix elements more severely is the
calculation of the polarization parameter $\alpha$. The primary reason is
$\alpha$ being a ratio of cross sections is less susceptible to theoretical
uncertainties from other sources, and we shall discuss this in section
\ref{section-numerical-analysis}. Concerning the matrix elements, we will 
incorporate the theoretical uncertainties, which have their origin in higher 
order corrections and the fit procedure. This we do by letting the 
matrix elements vary in ranges which cover almost all
values given in the literature. These ranges, are summarized in
appendix \ref{appendix-nrqcd} at the end of each table.
On the other hand, to avoid the predictions becoming too loose, we will 
impose three conditions on the values of the matrix elements.

\begin{enumerate}
\item All observables are restricted to physical values, i.e.,~for instance
  the momentum spectra, $d \Gamma/ d|\bvec{k}|$, have to be positive for
  all values of $|\bvec{k}|$.
\item The numerical values of $\langle {\cal O}_8^{\psi} (^{1} S_0)
  \rangle$ and  $\langle {\cal O}_8^{\psi} (^{3} P_0) \rangle$
  are required to give $M_r^\psi$ that lies within the determined
  range.
\item The resulting branching fractions, $Br(B \to \psi + X)$,
  have to match with experimental data \cite{Balest:1995jf,Groom:2000in} 
 within error bars. Hence, the variation of the matrix
  elements within their (rather large) errors is not
  independent anymore. Furthermore, $\varepsilon_P$, is constrained by this 
restriction, since its variation influences the branching ratio.
\end{enumerate}
The above conditions do allow us to constrain the theoretical
uncertainties to a reasonable measure.

In order to have a quantitative feeling for the influence of the $B$ 
cross section on the $J/\psi$ production cross section (figure
\ref{plot-error-cross-jpsi-B-input}) and the contributions of the individual
NRQCD matrix elements to the $J/\psi$ production cross section (figure
\ref{plot-contributions-cross-jpsi-nrqcd-separate}), we fix a set of input
matrix elements. Here, for these two figures, we have used the following
numerical matrix element values as input for calculating the $J/\psi$
production cross section and will refer to them as the {\em standard} set:
\begin{align}
\langle {\cal O}_1^{J/\psi} (^{3} S_1) \rangle &= 1.1~\textrm{GeV}^3~,
\nonumber \\
\langle {\cal O}_8^{J/\psi} (^{3} S_1) \rangle &= 0.0075~\textrm{GeV}^3~,
\label{inputv} \\
\langle {\cal O}_8^{J/\psi} (^{1} S_0) \rangle &= 0.055~\textrm{GeV}^3~,
\nonumber\\
\langle {\cal O}_8^{J/\psi} (^{3} P_0) \rangle &= - 0.006~\textrm{GeV}^5~, \nonumber
\end{align}
along with $\varepsilon_P = 0.012$. These values correspond to
$M_r^{J/\psi} = 0.046~\textrm{GeV}^3$. We
remind that our choice in (\ref{inputv}) is different from that made in 
\cite{Kniehl:1999vf, Palmer:1997wv}, since we are using a generic 
representative value taken from the allowed range as shown in 
appendix \ref{appendix-nrqcd}. The values are required to reproduce the
experimental branching ratio, $Br(B \to J/\psi + X)= (0.80\pm 0.08)\%$ taken 
from \cite{Balest:1995jf}. We reiterate and clarify that our terminology 
{\it ``standard''} is to be understood only in the context of the values in 
(\ref{inputv}) which are used to reproduce figures 
\ref{plot-error-cross-jpsi-B-input} and
\ref{plot-contributions-cross-jpsi-nrqcd-separate}.

\subsection{Other input parameters}

\subsubsection{Quark masses}
The $b$ quark mass, which is not very accurately known,
does not appear in our calculations. This is because, by virtue of
having adopted the PM, the quark mass is replaced by the
corresponding hadron masses which are known to high accuracy 
\cite{Groom:2000in}. Nevertheless, an
effective $b$ quark mass, $\langle m_b \rangle$, can be defined in the
framework of the PM as has been done in section
\ref{parton-model}. Hence, a variation of $\varepsilon_P$, is in some sense 
equivalent to that of $m_b$.

The value of the $c$ quark mass is strongly correlated with the values of the
matrix elements. We have chosen $m_c = 1.5$~GeV and do not vary it 
independently from the matrix elements, because a variation of $m_c$ by 
$\pm 100$ MeV is mostly taken into account in the uncertainties of the 
matrix elements. So, the influence of the uncertainties is not analyzed
separately. This value is also used in most standard
publications that have performed fits of the matrix elements. The alternative
choice, $m_c = M_\psi/2$, is of course also appropriate in the case of 
$J/\psi$, since it is only slightly heavier than $3$~GeV. But this 
choice is unsuitable for the excited states like $\psi'$ and $\chi_{cJ}$. 
For the latter charmonium states, it
would severely reduce the branching fraction $Br(B \to \psi + X)$, if one does
not perform a new extraction of the matrix elements with $m_c \sim
1.8$~GeV. Another argument for the use of $m_c = 1.5$~GeV, is that it
is a short-distance parameter in the calculation whereas bound state
effects which are responsible for the mass differences of the various
charmonium states are long-distance effects and thus have to be
included in the matrix elements.

The light quark mass $m_f$ is in general set to zero in the numerical
calculations. To estimate its influence on the final results we let it
vary in the range $m_f = 0 \ldots 150$~MeV. This has no significant
impact on our results.

\subsubsection{Peterson parameter}
Along with the matrix elements, the distribution function parameter
$\varepsilon_P$, is varied such that the theoretical branching fraction
$Br(B \to \psi + X)$ coincides with the measured value within
$1\sigma$, hence, this couples $\varepsilon_P$ to the values of the matrix
elements. The standard value for $\varepsilon_P$ has been determined in 
fragmentation processes of high energy
$b$ quarks (using LO QCD calculations) and $\varepsilon_P = 0.006
\pm 0.002$ \cite{Chrin:1987yd}. On the other hand, a more recent LO extraction
has found a somewhat larger value of $\varepsilon_P = 0.0126$ 
\cite{Binnewies:1998vm}. In the case of semi-leptonic $B$ decays, relatively 
small values of $\varepsilon_P$ are preferred \cite{Jin:1994vc}, whereas, an
application to hadronic $B$ decays show better results for larger
values, $\varepsilon_P = 0.008 \ldots 0.012$ \cite{Palmer:1997wv}.

For our analysis, we let $\varepsilon_P$ vary between $0.004$ and $0.018$, 
which covers the whole spectrum of its allowed numerical value. We note that
the relation between the fragmentation and the distribution function
is exact only in an infinite momentum frame for the $B$ meson
\cite{Bareiss:1989my}. Further, the wide range of values for $\varepsilon_P$
also reflects the fact that we deal with a spectrum of $B$ momenta which 
certainly is not infinite. Additionally, the range of (preferred) values
for $\varepsilon_P$ could be due to the different momentum transfer
involved in semi-leptonic and hadronic $B$ decays. 
\subsubsection{Wilson coefficients}
As already known for many years now \cite{Bergstrom:1994vc}, the
color singlet Wilson coefficient $C_{[1]}$ is not
only small compared to the color octet coefficient $C_{[8]}$, but
also strongly dependent on the factorization
scale, $\mu$. For a particular choice of $\mu$ in the range, 
$m_c < \mu < m_b$, the singlet coefficient even vanishes thereby, completely 
switching off the color singlet
contribution to charmonium production in $B$ decays. This problem is
not fixed even at next-to-leading order, and more
so, it even gives rise to a negative, i.e.,~unphysical decay rate 
\cite{Bergstrom:1994vc,Beneke:1998ks}. Solutions to this problem have been 
proposed, e.g.,~an alternative combined expansion in $\alpha_s$ and 
$C_{[1]}/C_{[8]}$ which seems to stabilize the evolution, but this proposal 
lacks a solid theoretical basis \cite{Beneke:1998ks}.
We estimate the uncertainties due to the factorization scale
dependence of the Wilson coefficients by letting $\mu$ vary between
its {\it standard value} of $4.7$~GeV 
\cite{Fleming:1997pt,Palmer:1997wv,Kniehl:1999vf} to $3$~GeV, and study the 
influence on the differential cross section and the polarization parameter, 
$\alpha$. 

\subsubsection{Sundries}
\label{section-various-parameters}
Other numerical parameters that occur in our calculation are the Fermi
coupling constant $G_F$, and the CKM matrix elements $|V_{cf}|^2$. We
have taken the standard values \cite{Groom:2000in} which are
\begin{align}
  G_F &= 1.1664 \cdot 10^{-5}~\textrm{GeV}^{-2}~. \nonumber\\
  |V_{cb}| &= 0.0402 \pm 0.0019~.
\end{align}
If the light quark mass $m_f$ is neglected, one can simply sum over
the flavor $q_f = d$, $s$ and apply unitarity to obtain the
approximate relation
\begin{equation}
  |V_{cd}|^2 + |V_{cs}|^2 \approx 1~.
\end{equation}
For the $B$ mass we adopt $M_B = 5.279$~GeV, because the $B^{\pm}$ and
$B^0$, $\bar{B}^0$ dominate in the $b$ hadron admixture that is
encountered at the Tevatron. We take the average lifetime to be, $\tau_B =
(1.564\pm 0.014) \cdot 10^{-12}$~s.
Branching fractions, $Br(B \to \psi + X)$, that are required to constrain
the set of matrix elements are taken from \cite{Groom:2000in}:
\begin{align}
  Br(B \to J/\psi(\textrm{direct}) + X) &= (8.0 \pm 0.8)\cdot 10^{-3} ~.
\nonumber\\
  Br(B \to \psi' + X) &= (3.5 \pm 0.5) \cdot 10^{-3} ~.\nonumber\\
  Br(B \to \chi_{c1}(\textrm{direct}) + X) &= (3.7 \pm 0.7) \cdot 10^{-3}~.
\end{align}
The branching ratios for the feed-down channels are taken to be
\begin{align}
  Br(\psi' \to J/\psi + X) &= (55 \pm 5)\% ~.\nonumber\\
  Br(\chi_{c1} \to J/\psi + \gamma) &= (27.3 \pm 1.6)\%~.
\end{align}
To normalize the cross sections according to  data the charmonium
branching fractions to $\mu^+ \mu^-$ are required and we take 
them to be
\begin{align}
  Br(J/\psi \to \mu^+ \mu^-) &= (5.88 \pm 0.10)\% ~.\nonumber\\
  Br(\psi' \to \mu^+ \mu^-) &= (1.03 \pm 0.35)\%~.
\end{align}
\section{Numerical analysis and discussion}
\label{section-numerical-analysis}
In this section we will present the numerical results of our
calculations and compare them with available experimental data;
the unpolarized quarkonium production cross sections from $B$
decays \cite{Daniels:1997, Abe:1997jz} and the $\alpha$ parameter 
\cite{Affolder:2000nn}. This analysis also illustrates the dependence of our
numerical results on the various input parameters and the associated errors 
that are involved.

\subsection{$B$ production cross section}
\label{section-B-cross-errors}
\begin{figure}[htb]
  \begin{center}
    \epsfig{file=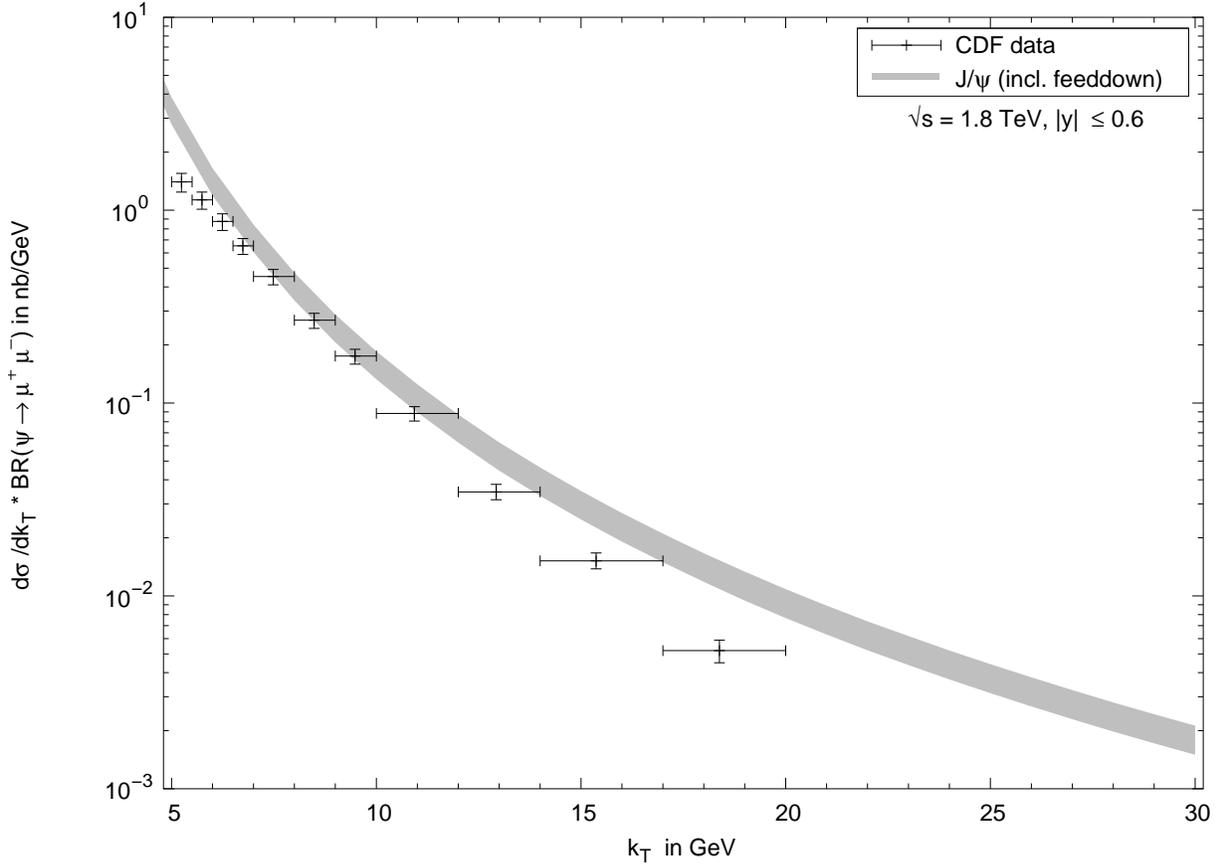,angle=-90,width=16cm}
  \end{center}
  \caption{Uncertainties in $d \sigma/ d k_T$ for $J/\psi$ production
    due to the experimental $B$ cross section.}
  \label{plot-error-cross-jpsi-B-input}
\end{figure}

The uncertainties of the $B$ cross section, $d \sigma/d p_T$,
have quite a large impact on the $J/\psi$ cross section and is of the order 
of $15 \ldots 20\%$. In figure
\ref{plot-error-cross-jpsi-B-input}, we show these uncertainties for 
unpolarized $J/\psi$ production cross sections when $d
\sigma/d p_T$ is varied between its lower and upper limit, as 
specified in table \ref{table-fit-B-production}. The matrix elements and 
Peterson parameter were adjusted to the standard values which are given in 
section \ref{section-analysis-strategy}.
For $J/\psi$ production, one clearly observes that the predicted cross section 
agrees quite well with the CDF data \cite{Abe:1997jz} for 
the range, $k_T = 7 \ldots 12$~GeV and is within the
statistical errors. For the rest of the $k_T$ values, it clearly
overestimates the data. Here, it is not a theoretical failure, but
rather an artifact of the fit procedure for the $B$ cross section. To
evaluate our formula in the $k_T$ range $4 \ldots 25$~GeV, we need to
extrapolate the $B$ cross section to $p_T$ values as low as $3$~GeV
and as high as $75$~GeV. Presently, data exist only for $p_T = 5 \ldots
20$~GeV. At the lower end $p_T \approx 5$~GeV, mass effects of the $b$
quark start to play an important role, causing the cross section
to stop rising as steeply as in the higher $p_T$ region. Also for very
large $p_T$, our fit seems to overestimate the data quite severely. In
the intermediate interval, where only $B$ mesons with transverse
momentum from the measured range contribute, we have a good agreement. 

However, as we focus on $\alpha$ which is a ratio 
of cross sections, it is almost unchanged by the variation of the input
momentum distribution for $b$ hadrons and we can extrapolate to very
small/large $p_T$. This makes the prediction for $\alpha$ insensitive to
the errors due to the fitting procedure. Numerically, $\alpha$ varies at 
the level of $10^{-3}$, hence the errors can be safely neglected.
The uncertainties of parameters that influence the differential cross section 
normalization have also been included in figure
\ref{plot-error-cross-jpsi-B-input}. Among these are the scale
factor $F_B$, whose error is specified in table
\ref{table-b-fractions}, the branching fraction, $Br(\psi \to \mu^+
\mu^-)$, and the CKM matrix element $V_{cb}$ which are given in section
\ref{section-various-parameters}.

The numerical impact of neglecting the rapidity dependence of the $B$
production cross section is noticeable at the level of
less than 3\% in the calculation of the $J/\psi$ cross section
and less than 1\% in the results for $\alpha$. The reason for this rather 
weak influence is that the
experimental cut on the $J/\psi$ rapidity ($|y| \le 0.6$) also imposes an
upper limit on $y_B$. A simple kinematic
calculation shows that the difference between $y$ and $y_B$ is
limited by the following expression
\begin{equation}
  |y_B - y| \le {\rm cosh^{-1}}{\left(\frac{\tfrac{1}{2}(M_B^2 +
   M_\psi^2 -m_f^2) - k_T p_T \cos{\phi}}{\sqrt{k_T^2 + M_\psi^2}
   \sqrt{p_T^2 + M_B^2}} \right)}~.
\end{equation}
For $\cos{\phi} = -1$ and $y = y_{\textrm{max}} = 0.6$, we find an
absolute upper limit on $y_B$ as a function of $k_T$ and $p_T$. In
our calculation this expression reaches its maximum for $k_T = 4$~GeV
and $p_T \approx 6$~GeV, giving a numerical value of $|y_B| \lesssim
0.95$ . As stated earlier, the dependence of the double differential $B$
cross section on $y_B$ is rather weak in this region and hence does
not lead to significant deviations if it is neglected. This justifies
our procedure in all posterity. 
\subsection{Dependence on matrix elements}
\label{matrixdepend}
Both the differential cross section and $\alpha$ are
strongly influenced by the variation of the matrix elements. For
$\alpha$, it is the main source of uncertainty, while all other errors, 
are canceled to a good accuracy, since $\alpha$ is a ratio of cross sections. 
We also remind that the variation of  $\varepsilon_P$ goes along with the 
variation of the matrix elements, because of the conditions that we imposed 
in section \ref{section-analysis-strategy}.

The direct $J/\psi$ production cross section shows a variation of $\pm
10\%$ on the matrix elements. It should be noted that $d \sigma/ d k_T$ is not
sensitive to the individual values of the matrix elements, but only to
the value of the resulting decay rate $\Gamma (B \to J/\psi + X)$
or, equivalently, to the branching fraction $Br(B \to J/\psi +
X)$. The maximum value of the cross section corresponds to a combination of
matrix elements and $\varepsilon_P$, leading to the upper limit value
of $Br(B \to J/\psi + X) = 0.88 \%$, whereas the minimum value
corresponds to the lower limit $Br(B \to J/\psi + X) = 0.72 \%$ (see
sections \ref{section-analysis-strategy} and
\ref{section-various-parameters}). 

\begin{figure}[htb]
  \begin{center}
    \epsfig{file=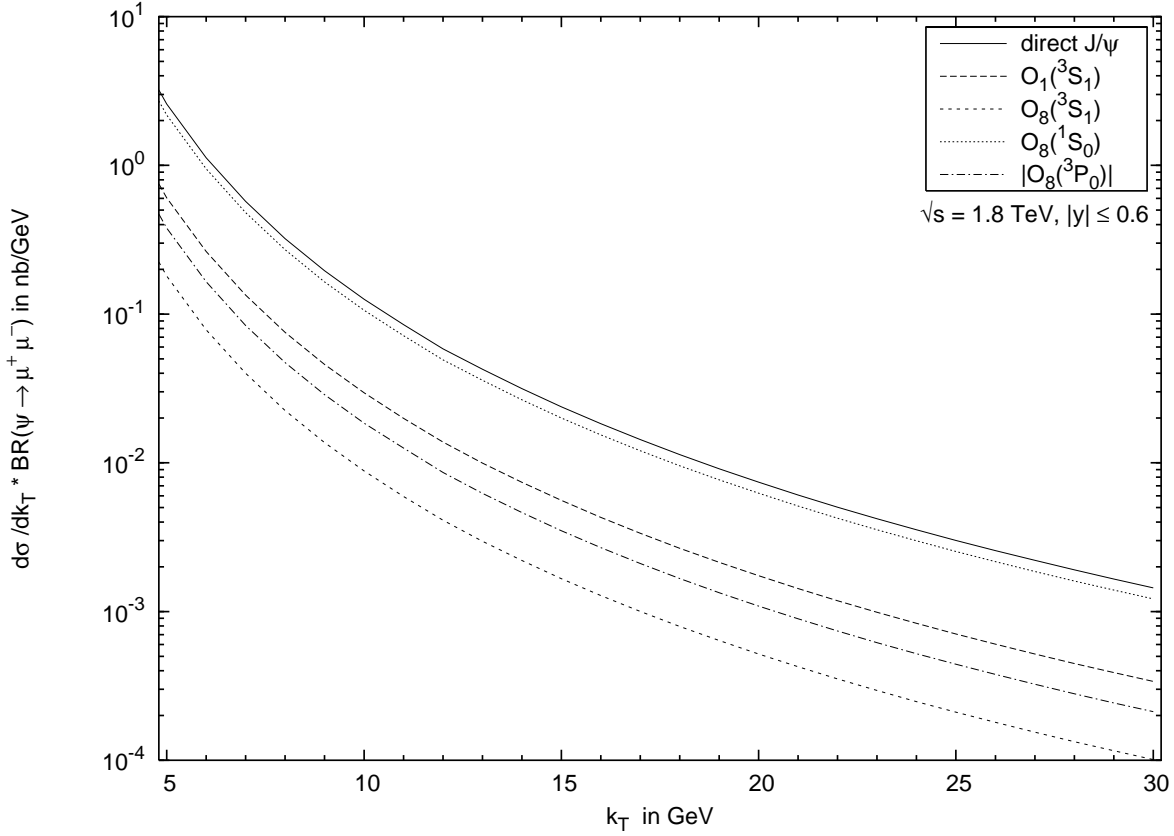,angle=-90,width=16cm}
  \end{center}
  \caption{Contributions of the different NRQCD matrix elements to the
    direct $J/\psi$ production cross section $d \sigma/ d k_T$.}
  \label{plot-contributions-cross-jpsi-nrqcd-separate}
\end{figure}

In figure \ref{plot-contributions-cross-jpsi-nrqcd-separate}, we show the
individual contributions of each matrix element to the unpolarized
direct $J/\psi$ production cross section, $d \sigma/ d k_T$, for a
standard set of matrix elements and $\varepsilon_P$ as given in section 
\ref{section-analysis-strategy}. It has to be
noted that for $\langle {\cal O}_8^{J/\psi}(^{3} P_0) \rangle$, we show
the absolute value of the contribution, since it is negative (see
section \ref{section-nrqcd-matrix-elements}).
As can be seen, the transverse momentum dependence to the different
contributions is approximately the same. Thus this feature does not allow us 
to concentrate on any one matrix element contribution in a particular
kinematic region. This has to be contrasted to prompt production, where the
contributions of the matrix elements have different $k_T$ dependence. 

To quantitatively illustrate $\alpha$'s dependence on the various matrix
elements, we give the helicity parameter for direct $J/\psi$
production, which we find to be 
\begin{equation}
  \alpha_{\textrm{direct}} = \frac{a_1 \langle {\cal O}_1^{J/\psi}(^{3} S_1) \rangle +
    a_8^{^3S_1} \langle {\cal O}_8^{J/\psi}(^{3} S_1) \rangle +
    a_8^{^3P_0} \langle {\cal O}_8^{J/\psi}(^{3} P_0) \rangle}
    {\langle {\cal O}_1^{J/\psi}(^{3} S_1) \rangle +
    b_8^{^3S_1} \langle {\cal O}_8^{J/\psi}(^{3} S_1) \rangle +
    b_8^{^3P_0} \langle {\cal O}_8^{J/\psi}(^{3} P_0) \rangle +
    b_8^{^1S_0} \langle {\cal O}_8^{J/\psi}(^{1} S_0) \rangle}~.
\label{element}
\end{equation}
To avoid the strong $k_T$ dependence of the cross section, we have
normalized the color singlet coefficient in the denominator to
unity. Note that the numerator in (\ref{element}) is independent of 
$\langle {\cal
  O}_8^{J/\psi}(^{1} S_0) \rangle$, because its contribution to the
decay rate is helicity independent and thus cancels out in the
numerator. The coefficients $a_n^{^{2s+1} L_J}$ and $b_n^{^{2s+1}
  L_J}$ are given in table \ref{table-prefactors-alpha} as functions
of $k_T$ for $\varepsilon_P = 0.012$~.
\begin{table}[htb]
  \begin{center}
    \begin{tabular}{|l|c|c|c|c|c|c|}
      \hline 
      $k_T$ in GeV & $a_1$ & $a_8^{^3S_1}$ & $a_8^{^3P_0}$ & $b_8^{^3S_1}$ &
      $b_8^{^3P_0}$ & $b_8^{^1S_0}$ \\
      \hline
       5 & $-0.178$ & $-7.78$ & $+12.8$ & $+43.7$ & $+103.8$ & $+67.9$ \\
       10 & $-0.151$ & $-6.58$ & $+11.3$ & $+43.7$ & $+105.0$ & $+68.1$ \\
       15 & $-0.145$ & $-6.33$ & $+11.1$ & $+43.7$ & $+105.5$ & $+68.3$ \\
       20 & $-0.142$ & $-6.22$ & $+11.0$ & $+43.7$ & $+105.6$ & $+68.3$ \\
       25 & $-0.142$ & $-6.20$ & $+10.9$ & $+43.7$ & $+105.6$ & $+68.4$ \\
      \hline
    \end{tabular}
  \end{center}
  \caption{Coefficients of the matrix elements that describe the short
      distance effects in $\alpha$ for $\varepsilon_P = 0.012$.}
  \label{table-prefactors-alpha}
\end{table}
Clearly, $\alpha$ is sensitive to the numerical values of the matrix
elements and also depends on the short-distance coefficients. Apart from
the normalization, due to identical short-distance coefficients, a variation 
of both the singlet and octet $^3 S_1$ matrix elements have a similar effect 
on $\alpha$, 
i.e.,~with an increase of the matrix
elements, $\alpha$ decreases (its absolute value increases). For
$\langle {\cal O}_8^{J/\psi}(^{1} S_0) \rangle$, the situation is
extremely simple, since it only contributes in the denominator,
hence an increase of its numerical value leads to a larger value of
$\alpha$ (reduction of the absolute value). A particularly strong
dependence is exhibited in the case of $\langle {\cal
  O}_8^{J/\psi}(^{3} P_0) \rangle$, as it can even change the
sign. In fact, for central values of the other matrix elements,
it is possible to find a reasonable value of $\langle {\cal
  O}_8^{J/\psi}(^{3} P_0) \rangle$ which leads to a diverging
polarization parameter. The reason being the short-distance
coefficient in the denominator in (\ref{element}) is so large that a 
small negative value of the matrix element can result in a vanishing
denominator. Obviously such a value of $\langle {\cal
  O}_8^{J/\psi}(^{3} P_0) \rangle$ violates the constraints of section
\ref{section-analysis-strategy}, because it corresponds to a negative
value of the cross section. Large positive values of the $^3 P_0$
matrix element are capable of creating a positive $\alpha$.

A notable feature which was observed in our numerical analysis is the
dependence of $\alpha$ on $\varepsilon_P$, which is very
pronounced; $|\alpha|$ decreases with increasing $\varepsilon_P$. This 
corresponds to a similar behavior of $\alpha$ with the $b$ quark mass as
observed in \cite{Fleming:1997pt}; the smaller $m_b$ (or larger 
$\varepsilon_P$) becomes, the smaller $|\alpha|$ gets to be. However, at the 
Tevatron, since the shape of the distribution function itself is not too 
important, this behavior is mainly related to determining the effective $b$ 
quark mass, as introduced in (\ref{eq-effective-b-mass}).

As stated earlier, in our analysis, the various matrix elements and 
$\varepsilon_P$ are not completely independent, but their values are 
correlated because of the conditions imposed in section
\ref{section-analysis-strategy}. Hence, the extreme values 
for $\alpha$ are a {\it compromise} of the tendencies described
above. The maximum of $\alpha$ corresponds to values of $\langle
{\cal O}_1^{J/\psi}(^{3} S_1) \rangle$ and $\langle {\cal
  O}_8^{J/\psi}(^{3} S_1) \rangle$ which are close to their lower
limits as specified in appendix \ref{appendix-nrqcd}. $\langle {\cal
  O}_8^{J/\psi}(^{1} S_0) \rangle$ is also rather small, $\langle {\cal
  O}_8^{J/\psi}(^{3} P_0) \rangle$ is large and positive and
$\varepsilon_P$ is at its allowed maximum. The minimum of $\alpha$ is
reached for ${\cal O}_1^{J/\psi}(^{3} S_1) \rangle$ and $\langle {\cal
  O}_8^{J/\psi}(^{3} S_1) \rangle$ being close to their upper limits,
$\langle {\cal O}_8^{J/\psi}(^{1} S_0) \rangle$ at an intermediate
value and $\langle {\cal O}_8^{J/\psi}(^{3} P_0) \rangle$ being close
to zero, but negative. $\varepsilon_P$ is at its lower bound, thus
illustrating the strong dependence of $\alpha$ on this parameter.

\subsection{Other theoretical uncertainties}
The variation of the factorization scale,
$\mu$, between its standard value $m_b = 4.7$~GeV and ${2 m_c
= 3}$~GeV changes the results significantly. The obvious reason
for this behavior is that lowering $\mu$ from $4.7$~GeV to
$3$~GeV is equivalent to reducing the color singlet
matrix elements by $\sim 58\%$ and at the same time increasing all
octet matrix elements by $\sim 6\%$ as can be estimated from figure
\ref{plot-wilson-coefficients}. The net effect is a $10 \%$
reduction of the unpolarized cross section, $d \sigma/d k_T$, and an
increase of $20 \ldots 25 \%$ in $\alpha$.

Increasing the light quark mass $m_f$, $q_f = d$, $s$, from $0$ to
$150$~MeV causes a reduction of the unpolarized cross section of
approximately $1.5\%$ if all other parameters are left
unchanged. Again $\alpha$
is less strongly affected and increases by less than $1\%$~.
The uncertainties of the remaining quantities exclusively affect the
cross section, because they have only an influence on its
normalization and thus do not affect the predictions for $\alpha$. To this 
category, belong the CKM matrix element $|V_{cb}|$, the average $B$ life time
$\tau_B$ and the branching fraction $Br( \psi \to \mu^+ \mu^-)$ which
is used to normalize the spectrum. In the combined error of $d
\sigma/d k_T$, these uncertainties have been included.
\begin{figure}[htb]
  \begin{center}
    \epsfig{file=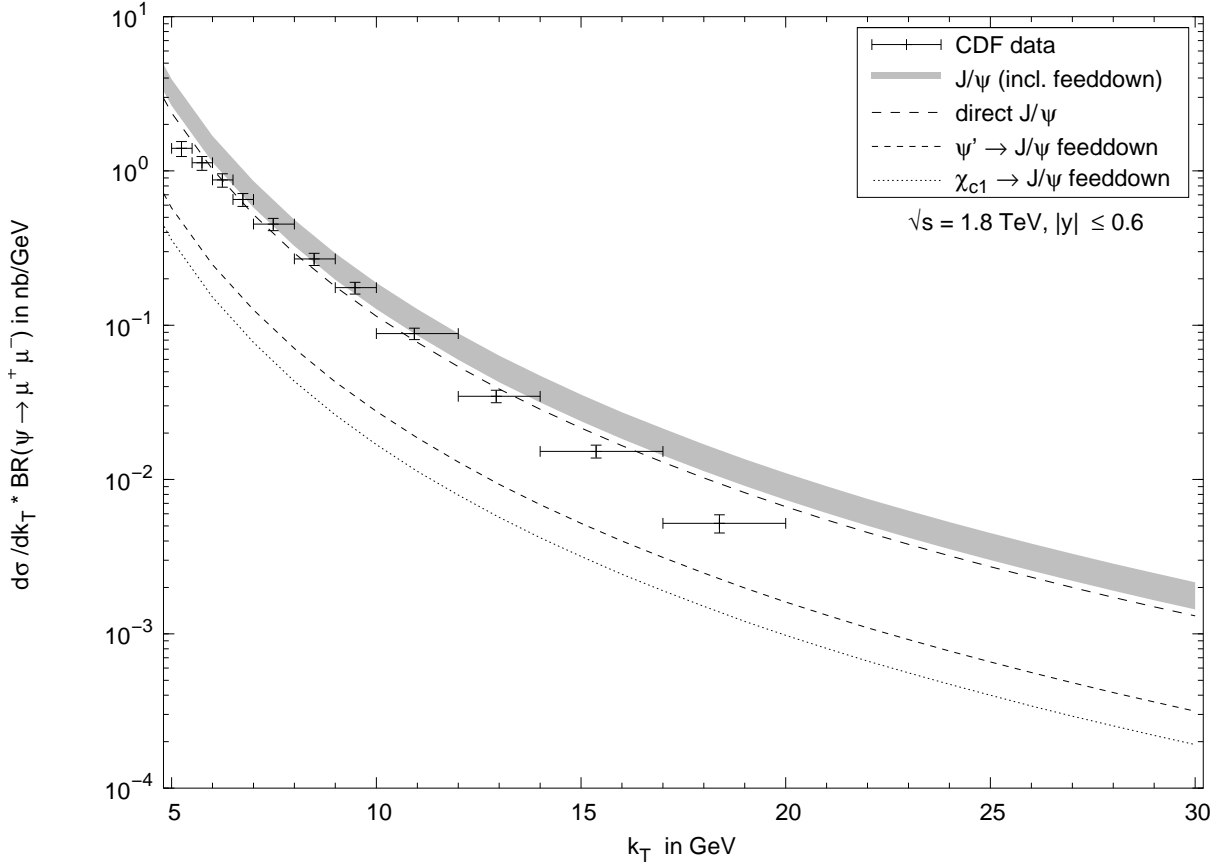,angle=-90,width=16cm}
  \end{center}
  \caption{Unpolarized cross section $d \sigma/ d k_T$ for $J/\psi$
    production with contribution from all different channels and the
    combined error.}
  \label{plot-error-cross-jpsi-allcombined}
\end{figure}
\begin{figure}[htb]
  \begin{center}
    \epsfig{file=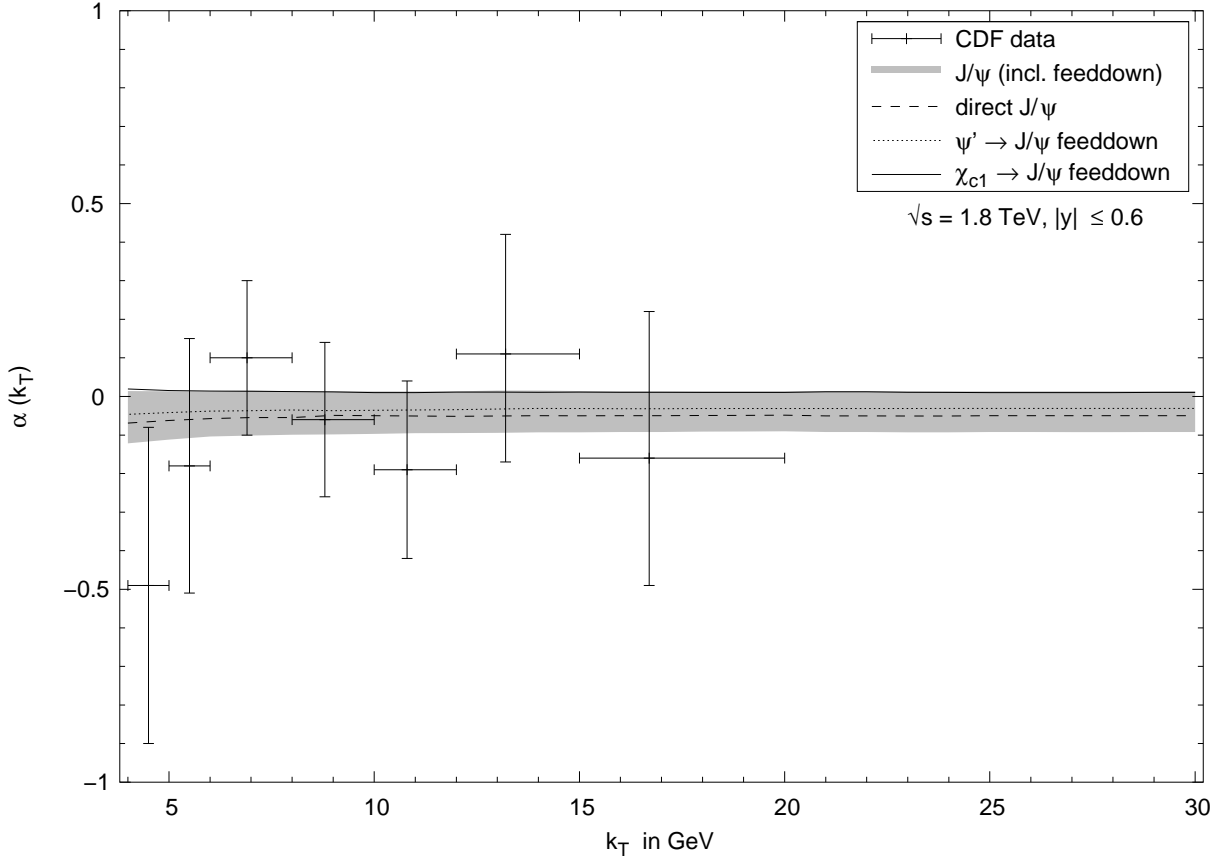,angle=-90,width=16cm}
  \end{center}
  \caption{$\alpha$ for $J/\psi$ from $B$ decays with contributions from all
    different channels and the combined error.}
  \label{plot-error-alpha-jpsi-allcombined}
\end{figure}
\begin{figure}[htb]
  \begin{center}
    \epsfig{file=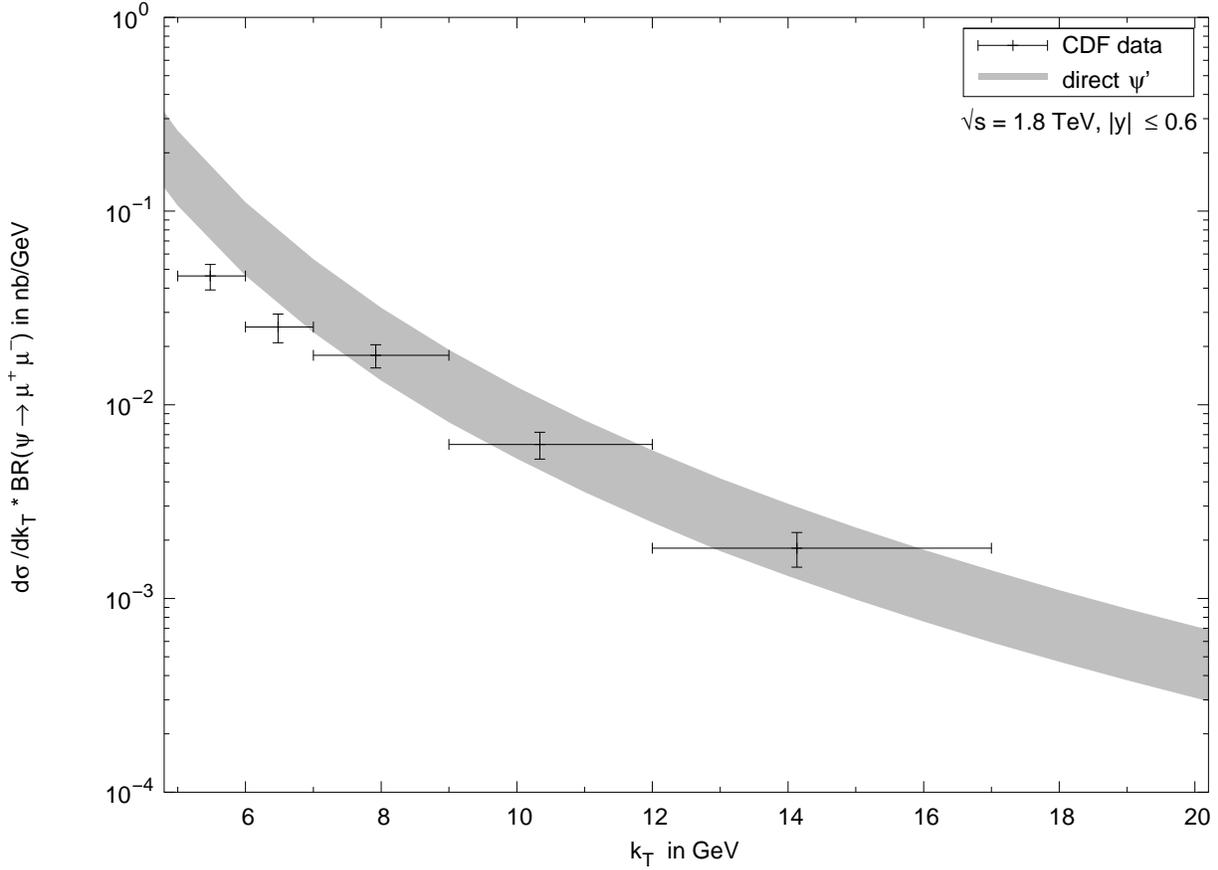,angle=-90,width=16cm}
  \end{center}
  \caption{Unpolarized cross section $d \sigma/ d k_T$ for $\psi'$
    production with the combined error.}
  \label{plot-error-cross-psi-prime-allcombined}
\end{figure}
\begin{figure}[htb]
  \begin{center}
    \epsfig{file=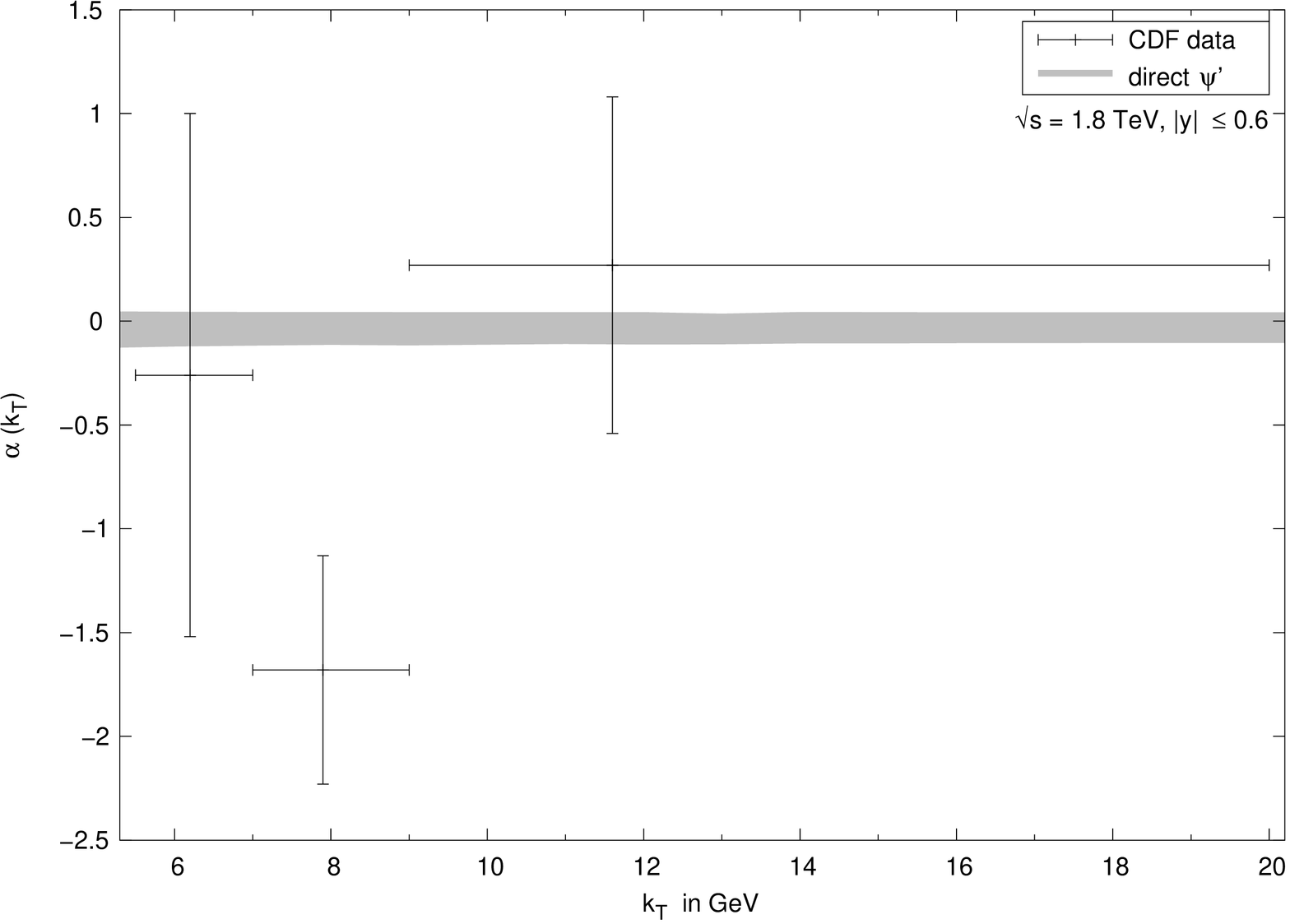,angle=-90,width=16cm}
  \end{center}
  \caption{$\alpha$ for $\psi'$ from $B$ decays with the combined error.}
  \label{plot-error-alpha-psi-prime-allcombined}
\end{figure}
\subsection{Final results}
The final results for $J/\psi$ as well as the experimental data are
presented in figures \ref{plot-error-cross-jpsi-allcombined} and
\ref{plot-error-alpha-jpsi-allcombined}. The corresponding quantities
for $\psi'$, to which no feed-down channels contribute, are shown in
figures \ref{plot-error-cross-psi-prime-allcombined} and
\ref{plot-error-alpha-psi-prime-allcombined}.

The unpolarized $J/\psi$ cross section shows a good agreement with the
experimental data \cite{Abe:1997jz} within error bars at intermediate
transverse momenta $k_T$. On the other hand it overestimates the data
quite strongly for small and large $k_T$, a feature which is a result
of extrapolating the $B$ cross section (see section
\ref{section-B-cross-errors} for details).

Our prediction for the polarization parameter $\alpha$ is consistent
with $0$, but the central value
prefers to be small and negative. For the higher transverse momentum
range $k_T = 10 \ldots 30$~GeV, we have $\alpha = -0.04 \pm 0.06$. We
find a good agreement with
experimental data \cite{Affolder:2000nn}, although one has to admit
that the statistical experimental uncertainties are enormous and for
some $k_T$ bins cover as much as one third of the theoretically
allowed parameter space of $\alpha$. The prediction for $\alpha$ is
almost independent of $k_T$ and only shows a slight decrease towards
the lowest transverse momentum values displayed in figure
\ref{plot-error-alpha-jpsi-allcombined}. Also the central values of
the experimental data exhibit such a tendency, but the decrease at low
$k_T$ is much more significant. Even here the predictions lie within 
$1\sigma$ deviation of the experimental data.

Obviously on the basis of the current data it is not possible to constrain 
the values of the poorly determined color octet $^1 S_0$ and $^3 P_0$
matrix elements any further. Nevertheless, there
are good prospects that the situation on the experimental side
will improve in the near future. Run II of the Tevatron is in progress for
one and a half years already and the $B$ physics group of the CDF
collaboration expects to increase statistics effectively by a factor
of 50 \cite{Anikeev:2001rk}. This will reduce the
statistical error at least in the low and medium transverse
momentum range significantly and also data in higher $k_T$ bins will
become available. Hence it is expected that after Run II only systematical
errors will dominate the uncertainties of the $\alpha$ measurement
\cite{Anikeev:2001rk}. Even if those will not be improved, $\alpha$
will have an error of the order of $\sim 0.02$, enough to exclude a
good part of the numerical ranges of the matrix elements.

Since our calculation can be equally applied to $\psi'$ production
without significant modifications, we
have extended the analysis to this charmonium state. For intermediate
and large values of $k_T$ the unpolarized cross section in figure
\ref{plot-error-cross-psi-prime-allcombined} agrees with the CDF data
within error bars. The reason for the excess at low $k_T$ is the same
as in the case of $J/\psi$ discussed earlier. With $\alpha = -0.03 \pm 0.07$ 
for $k_T =10 \ldots 30$~GeV, the polarization parameter for $\psi'$ does
not differ significantly from the one for $J/\psi$, but here a
comparison with data is almost impossible as can be seen from figure
\ref{plot-error-alpha-psi-prime-allcombined}. The situation is
worse than for $J/\psi$, because the error bars are huge and one of the
data points even lies outside the allowed region for $\alpha$, which is
restricted to the interval $[-1, +1]$ for theoretical reasons. On the
other hand, once more precise data will be available, to derive
tighter constraints on the $\psi'$ matrix elements will be more
straight forward than for $J/\psi$, because all $\psi'$ are directly
produced from $B$ decays, since no feed-down channels are known to contribute. 
\section{Summary and conclusions}
\label{summary}
In this paper, using the NRQCD formalism and the PM,
we have calculated the semi-inclusive decay rate,
$\Gamma(B \to J/\psi + X)$ with 
polarized $J/\psi$ as final states. Subsequently these results were
generalized to the case of other $J = 1$ quarkonium states, 
i.e.,~$\psi'$ and $\chi_{c1}$. The results were applied to the Fermilab 
Tevatron setup. In this case, we calculated
the differential cross sections for unpolarized $J/\psi$ and $\psi'$
production in $B$ decays and, the polarization parameter $\alpha$ for 
$J/\psi$ and $\psi'$,
originating from $B$ decays. The $B$ meson production cross section in
$p \bar{p}$ collisions at the Tevatron was implemented by a phenomenological
fit to the CDF data \cite{Affolder:2000nn}. We 
considered the feed-down from $\psi'$ and
$\chi_{c1}$ for $J/\psi$ production. To obtain a meaningful comparison
of our predictions with experimental data, we
carried out a detailed analysis of the various theoretical uncertainties 
involved. In particular, it was shown that $\alpha$ is almost not
influenced by most input parameters, except for the 
matrix elements and the distribution function parameter,
$\varepsilon_P$. Therefore, for an extraction of this parameter, it is 
pertinent for more precise numerical values of the matrix elements, especially,
the poorly determined $\langle {\cal O}_8^{\psi} (^{1} S_0) \rangle$
and $\langle {\cal O}_8^{\psi} (^{3} P_0) \rangle$.

To perform such an improved fit of these non-perturbative matrix
elements, the precision of the data has to be increased
significantly which is also expected in Run II of the Tevatron. 
Furthermore, it would also be desirable to experimentally separate direct
$J/\psi$ production from the feed-down channels. This restricts the
uncertainties, simply because the number of relevant matrix
elements get reduced . Theoretically, an inclusion of higher order
corrections would be preferable to reduce the errors due to factorization 
scale dependence, in particular that of the color
singlet Wilson coefficient. Since it is known that this cannot be
achieved at next-to-leading order, a next-to-next-to-leading order
calculation might be necessary \cite{Beneke:1998ks}. Also, a better
knowledge of $\varepsilon_P$
would improve the precision of $\alpha$. For $\psi'$ this could be
accomplished with a fit to the unpolarized momentum spectrum of $\psi'$
in $B$ decays to more accurate CLEO \cite{Anderson:2002jf} and BaBar
\cite{Aubert:2002yk} data
that have become available very recently. The feed-down channels and
resonant two body final states ($J/\psi + K$ and $J/\psi + K^\star$)
complicate such a fit for $B \to J/\psi + X$.
Additionally, it might be worth to try a different parameterization for the
heavy quark distribution function, because in $b$ fragmentation, which
serves as a motivation for the distribution function, the Peterson
form might not be appropriate \cite{Cacciari:2002pa}.

A significant reduction of the matrix element errors is mostly likely
to be achieved with the help of a global fit. Among the processes that
could contribute to this fit, $\alpha$ from $B$ decays might play an
important role, being one of the few quantities that is sensitive to
the individual values of the $\langle {\cal O}_8^{\psi} (^{1} S_0) \rangle$
and $\langle {\cal O}_8^{\psi} (^{3} P_0) \rangle$ matrix elements.
\vspace{0.5cm}

\centerline{\bf Acknowledgments}  
\noindent
This work has been supported by the
Bundesministerium f\"ur Bildung, Wissenschaft, Forschung und Technologie,
Bonn under contract no.~05HT1PEA9. We wish to thank E.A. Paschos
for suggesting this problem and for many useful discussions.
\begin{appendix}
\section{Appendix}
\subsection{NRQCD Matrix Elements}
\label{appendix-nrqcd}
Here, we present numerical values of the matrix elements from the
literature that are required in our analysis. The tables below are structured
as follows. Apart from the numerical value, including statistical as well as
systematical errors (wherever given) we refer to the method/process that has 
been used to extract them and the corresponding references to the original
publications. This list only
provides an overview of the numerical values that we have considered in our
analysis and is by no means meant to be exhaustive. 
In the last line of each table we give a range for the corresponding
matrix element that we have used in our calculation as described in
section \ref{section-analysis-strategy}.
\subsubsection{$J/\psi$ Matrix Elements}
\addcontentsline{toc}{subsection}{\numberline{}$J/\psi$ Matrix Elements}
\begin{center}
  \begin{tabular}{|p{5cm}|p{8cm}|p{1.74cm}|}
    \hline
    $\vphantom{\Big( \Big)} \langle {\cal O}_1^{J/\psi} (^{3} S_1) \rangle$ in GeV$^3$ &
    method/process & Reference \\
    \hline
    $0.994 \pm 0.002 ^{+0.204} _{-0.145}$ & lattice calculation & \cite{Bodwin:1996tg} \\
    $1.16 $ & Buchm\"uller-Tye potential & \cite{Eichten:1995ch} \\
    $1.1 \pm 0.1$ & $J/\psi \to e^+ e^-$ & \cite{Groom:2000in} \\
    $0.763 \pm 0.054$ & $J/\psi \to e^+ e^-$ (without QCD) &
    \cite{Kniehl:1998qy} \\
    $ 1.3 \pm 0.1$ & $J/\psi \to e^+ e^-$ (incl.~LO QCD) & \cite{Kniehl:1998qy,
      Braaten:1999qk} \\
    \hline
    $1.1 \pm 0.1$ & & \\
    \hline
  \end{tabular}
\end{center}
\begin{center}
  \begin{tabular}{|p{5cm}|p{8cm}|p{1.74cm}|}
    \hline
    $\vphantom{\Big( \Big)}\langle {\cal O}_8^{J/\psi} (^{3} S_1) \rangle$ in $10^{-3}$~GeV$^3$ &
    method/process & Reference \\
    \hline
    $6.6 \pm 2.1$ & hadroproduction (CDF data) & \cite{Cho:1996vh, Cho:1996ce} \\
    $3.94 \pm 0.63$ & hadroproduction (CDF data) & \cite{Kniehl:1998qy} \\
    $10.6 \pm 1.4 ^{+10.5}_{-5.9}$ & hadroproduction (CDF data) &
    \cite{Beneke:1997yw} \\
    $9.6 \pm 1.5$ & hadroproduction (CDF data) & \cite{Sanchis-Lozano:1999um} \\
    $4.4 \pm 0.7$ & hadroproduction (CDF data) & \cite{Braaten:1999qk} \\
    \hline
    $6.0\pm3.0$ & & \\
    \hline
  \end{tabular}
\end{center}
\begin{center}
  \begin{tabular}{|p{5cm}|p{0.87cm}|p{6.7cm}|p{1.74cm}|}
    \hline
    $\vphantom{\Big( \Big)}M_r^{J/\psi}$ in $10^{-2}$~GeV$^3$ & $r$ &
    method/process & Reference \\
    \hline
    $6.6 \pm 1.5$ & $3$ & hadroproduction (CDF data) & \cite{Cho:1996vh, Cho:1996ce} \\
    $6.52 \pm 0.67$ & $3.47$ & hadroproduction (CDF data) & \cite{Kniehl:1998qy} \\
    $3.0$ & $7$ & hadroproduction (fixed target) & \cite{Beneke:1996tk}\\
    $8.7 \pm 0.9$ & $3.4$ & hadroproduction (CDF data) & \cite{Braaten:1999qk}
    \\
    $2.0$ & $7$ & photoproduction & \cite{Fleming:1997jx} \\
    $4.38 \pm 1.15 ^{+1.52}_{-0.74}$ & $3.5$ & hadroproduction (CDF data) &
    \cite{Beneke:1997yw} \\
    $1.5$ & $3.1$ & $Br(B \to J/\psi + X)$ (CLEO data) & \cite{Beneke:1998ks}\\
    $1.32 \pm 0.21$ & 3 & hadroproduction (CDF data) &
    \cite{Sanchis-Lozano:1999um} \\
    \hline
    $5.0 \pm 4.0 $ & $3.5$ & & \\
    \hline
  \end{tabular}
\end{center}
\begin{center}
  \begin{tabular}{|p{5cm}|p{8cm}|p{1.74cm}|}
    \hline
    $\vphantom{\Big( \Big)}\langle {\cal O}_8^{J/\psi} (^{1} S_0) \rangle$ in $10^{-2}$~GeV$^3$ &
    method/process & Reference \\
    \hline
    $4.0$ & leptoproduction & \cite{Fleming:1997jx} \\
    $14.5$ & $Br(B \to J/\psi + X)$ (CLEO data) & \cite{Kniehl:1999vf} \\
    \hline
    $6.5 \pm 5.5$ & & \\
    \hline
  \end{tabular}
\end{center}
\begin{center}
  \begin{tabular}{|p{5cm}|p{8cm}|p{1.74cm}|}
    \hline
    $\vphantom{\Big( \Big)}\langle {\cal O}_8^{J/\psi} (^{3} P_0) \rangle$ in $10^{-2}$~GeV$^5$ &
    method/process & Reference \\
    \hline
    $-0.3 \cdot m_c^2$ & leptoproduction & \cite{Fleming:1997jx} \\
    $-5.51$ & $Br(B \to J/\psi + X)$ (CLEO data) & \cite{Kniehl:1999vf} \\
    \hline
    $-4.0 \pm 6.0$ & & \\
    \hline
  \end{tabular}
\end{center}
\subsubsection{$\psi'$ Matrix Elements}
\addcontentsline{toc}{subsection}{\numberline{}$\psi'$ Matrix Elements}
\begin{center}
  \begin{tabular}{|p{5cm}|p{8cm}|p{1.74cm}|}
    \hline
    $\vphantom{\Big( \Big)}\langle {\cal O}_1^{\psi'} (^{3} S_1) \rangle$ in GeV$^3$ &
    method/process & Reference \\
    \hline
    $0.440 \pm 0.043$ & $\psi' \to e^+ e^-$ (incl.~LO QCD) &
    \cite{Kniehl:1999vf} \\
    $ 0.65 \pm 0.6$ & $\psi' \to e^+ e^-$ (incl.~NLO QCD) & \cite{Braaten:1999qk}\\
    $0.76$ & Buchm\"uller-Tye potential & \cite{Eichten:1995ch} \\
    $0.11$ & leptonic decay rate & \cite{Braaten:1995vv} \\
    \hline
    $0.6 \pm 0.2$ & & \\
    \hline
  \end{tabular}
\end{center}
\begin{center}
  \begin{tabular}{|p{5cm}|p{8cm}|p{1.74cm}|}
    \hline
    $\vphantom{\Big( \Big)}\langle {\cal O}_8^{\psi'} (^{3} S_1) \rangle$ in $10^{-3}$ GeV$^3$ &
    method/process & Reference \\
    \hline
    $6.20 \pm 0.95$ & hadroproduction (CDF data) & \cite{Kniehl:1999vf} \\
    $4.2$ & hadroproduction (CDF data)& \cite{Braaten:1995vv} \\
    $4.6 \pm 1.0$ & hadroproduction (CDF data) & \cite{Cho:1996vh, Cho:1996ce}
    \\
    $4.4 \pm 0.8 ^{+4.3} _{-2.4}$ & hadroproduction (CDF data) &
    \cite{Beneke:1997yw} \\
    $4.2 \pm 1.0$ & hadroproduction (CDF data) & \cite{Braaten:1999qk} \\
    \hline
    $5.5 \pm 2.5$ & & \\
    \hline
  \end{tabular}
\end{center}
\begin{center}
  \begin{tabular}{|p{5cm}|p{0.87cm}|p{6.7cm}|p{1.74cm}|}
    \hline
    $\vphantom{\Big( \Big)}M_r^{\psi'}$ in $10^{-2}$~GeV$^3$ & $r$ &
    method/process & Reference \\
    \hline
    $1.8 \pm 0.6$ & $3$ & hadroproduction (CDF data) & \cite{Cho:1996vh, Cho:1996ce} \\
    $1.79 \pm 0.51$ & $2.56$ & hadroproduction (CDF data) & \cite{Kniehl:1999vf} \\
    $1.8 \pm 0.56 ^{+0.62} _{-0.30}$ & $3.5$ & hadroproduction & \cite{Beneke:1997yw}\\
    $0.52$ & $7$ & hadroproduction (fixed target) & \cite{Beneke:1996tk}\\
    $1.3 \pm 0.5$ & $3.5$ & hadroproduction (CDF data) & \cite{Braaten:1999qk} \\
    $0.6$ & $3.1$ & $Br(B \to \psi' + X)$ (CLEO data) & \cite{Beneke:1998ks}\\
    \hline
    $1.75 \pm 1.25$ & $3.0$ & & \\
    \hline
  \end{tabular}
\end{center}
\begin{center}
  \begin{tabular}{|p{5cm}|p{8cm}|p{1.74cm}|}
    \hline
    $\vphantom{\Big( \Big)}\langle {\cal O}_8^{\psi'} (^{1} S_0) \rangle$ in $10^{-2}$~GeV$^3$ &
    method/process & Reference \\
    \hline
    $-0.96$ & $Br(B \to J/\psi + X)$ (CLEO data) & \cite{Kniehl:1999vf} \\
    \hline
    $2.5 \pm 2.5$ & & \\
    \hline
  \end{tabular}
\end{center}
\begin{center}
  \begin{tabular}{|p{5cm}|p{8cm}|p{1.74cm}|}
    \hline
    $\vphantom{\Big( \Big)}\langle {\cal O}_8^{\psi'} (^{3} P_0) \rangle$ in $10^{-2}$~GeV$^5$ &
    method/process & Reference \\
    \hline
    $2.58$ & $Br(B \to J/\psi + X)$ (CLEO data) & \cite{Kniehl:1999vf} \\
    \hline
    $0.0 \pm 10.0$ & & \\
    \hline
  \end{tabular}
\end{center}
\subsubsection{$\chi_{cJ}$ Matrix Elements}
\addcontentsline{toc}{subsection}{\numberline{}$\chi_{cJ}$ Matrix Elements}
\begin{center}
  \begin{tabular}{|p{5cm}|p{8cm}|p{1.74cm}|}
    \hline
    $\vphantom{\Big( \Big)}\langle {\cal O}_1^{\chi_{c0}} (^{3} P_0) \rangle$ in $10^{-2}$~GeV$^5$ &
    method/process & Reference \\
    \hline
    $4.8 \cdot m_c^2$ & Buchm\"uller-Tye potential & \cite{Eichten:1995ch} \\
    $4.4 \cdot m_c^2$ & Buchm\"uller-Tye potential & \cite{Eichten:1995ch, Beneke:1996tk} \\
    $22.9 \pm 2.5$ & hadroproduction (CDF data) & \cite{Kniehl:1998qy} \\
    $8.8 \pm 2.13$ & hadronic $\chi_{cJ}$ decays & \cite{Bodwin:1992qr} \\
    $8.9 \pm 1.3$ & $\chi_{c2} \to \gamma \gamma$ & \cite{Braaten:1999qk} \\
    \hline
    $6.0 \pm 4.0$ & & \\
    \hline
  \end{tabular}
\end{center}
\begin{center}
  \begin{tabular}{|p{5cm}|p{8cm}|p{1.74cm}|}
    \hline
    $\vphantom{\Big( \Big)}\langle {\cal O}_8^{\chi_{c0}} (^{3} S_1) \rangle$ in $10^{-3}$ GeV$^3$ &
    method/process & Reference \\
    \hline
    $4.5 \ldots 6.5$ & $B \to \chi_{c2} + X$ & \cite{Beneke:1998ks} \\
    $0.681 \pm 0.175$ & hadroproduction (CDF data) & \cite{Kniehl:1998qy} \\
    $3.2 \pm 1.4$ & $\Gamma( B \to \chi_{c2} + X) / \Gamma( B \to e \nu_e +X)$
    & \cite{Braaten:1995vv} \\
    $1.39 \pm 0.17$ & hadroproduction (CDF data) & \cite{Kniehl:1999vf} \\
    $2.3 \pm 0.3$ & hadroproduction (CDF data) & \cite{Braaten:1999qk} \\
    \hline
    $3.5 \pm 3.0$ & & \\
    \hline
  \end{tabular}
\end{center}
\end{appendix}

\end{document}